\documentclass[preprint,12pt,sort&compress]{elsarticle}

\usepackage[margin=1in]{geometry}
\usepackage{helvet}

\usepackage{amsmath, graphicx, tikz, float}
\usetikzlibrary{arrows.meta, positioning, calc}
\usepackage{setspace}
\usepackage{titlesec}

\titleformat{\section}{\normalfont\large\bfseries}{\thesection}{0.5em}{}
\titleformat{\subsection}{\normalfont\large\bfseries}{\thesubsection}{0.5em}{}
\titleformat{\subsubsection}{\normalfont\normalsize\bfseries}{\thesubsubsection}{0.5em}{}
\titlespacing*{\section}{0pt}{1.2ex}{0.8ex}
\titlespacing*{\subsection}{0pt}{1.0ex}{0.6ex}
\titlespacing*{\subsubsection}{0pt}{0.8ex}{0.5ex}

\newcommand{\sectionstylee}[1]{\noindent{\normalfont\normalsize\sffamily\bfseries #1}}

\makeatletter
\def\ps@pprintTitle{%
 \let\@oddhead\@empty
 \let\@evenhead\@empty
 \let\@oddfoot\@empty
 \let\@evenfoot\@oddfoot}
\makeatother

\begin{document}

\title{Biomechanically Informed Image Registration for Patient-Specific Aortic Valve Strain Analysis}

\author[1,2]{Mohsen Nakhaei\corref{cor1}}
\ead{nakhaeim@chop.edu}
\author[2]{Alison M. Pouch}
\author[2]{Silvani Amin}
\author[1]{Matthew Daemer}
\author[1]{Christian Herz}
\author[2]{Natalie Yushkevich}
\author[3]{Lourdes Al Ghofaily}
\author[4]{Nimesh Desai}
\author[5]{Joseph Bavaria}
\author[1]{Matthew A. Jolley}
\author[1,6,7]{Wensi Wu\corref{cor1}}
\ead{wensiwu@seas.upenn.edu}

\cortext[cor1]{Corresponding authors}

\affiliation[1]{organization={Department of Anesthesiology and Critical Care Medicine, Children’s Hospital of Philadelphia}, city={Philadelphia}, state={PA}, country={USA}}
\affiliation[2]{organization={Department of Radiology and Bioengineering, University of Pennsylvania}, city={Philadelphia}, state={PA}, country={USA}}
\affiliation[3]{organization={Department of Anesthesiology, University of Pennsylvania}, city={Philadelphia}, state={PA}, country={USA}}
\affiliation[4]{organization={Department of Surgery, University of Pennsylvania}, city={Philadelphia}, state={PA}, country={USA}}
\affiliation[5]{organization={Department of Cardiac Surgery, Jefferson Health}, city={Philadelphia}, state={PA}, country={USA}}
\affiliation[6]{organization={Department of Mechanical Engineering and Applied Mechanics, University of Pennsylvania}, city={Philadelphia}, state={PA}, country={USA}}
\affiliation[7]{organization={Cardiovascular Institute, Children's Hospital of Philadelphia}, city={Philadelphia}, state={PA}, country={USA}}

\begin{abstract}
\sectionstylee{Purpose} Aortic valve (AV) biomechanics play a critical role in maintaining normal cardiac function. Pathological variations, particularly in bicuspid aortic valves, alter leaflet loading, increase strain, and accelerate disease progression. Accurate patient-specific characterization of valve geometry and deformation is therefore essential for predicting disease progression and guiding durable repair. However, existing imaging and computational methods often fail to capture rapid valve motion and complex patient-specific features, limiting precise biomechanical assessment. \\
\sectionstylee{Methods} To address these limitations, we developed an image registration framework coupled with the finite element method (FEM) to improve AV tracking and biomechanical evaluation. Patient-specific valve geometries derived from 4D echocardiography and CT were used to simulate AV closure and generate intermediate deformation states. These FEM-generated states facilitated leaflet tracking, while image registration corrected misalignment between simulations and imaging data.\\
\sectionstylee{Results} In 20 patients, FEM-augmented registration improved tracking accuracy by 40\% compared with direct registration. This improvement enabled more reliable strain estimation by measuring leaflet deformation directly from imaging and reducing uncertainties associated with boundary conditions and material assumptions. Using the improved tracking results, areal, Green-Lagrange, and deviatoric strains were quantified in adult trileaflet and bicuspid valves, as well as pediatric patients, revealing distinct deformation patterns across valve groups. Convergence in mean deviatoric strain between adult trileaflet and pediatric valves suggests volumetric deformation underlies age- and size-related differences in AV mechanics.\\
\sectionstylee{Conclusion} Overall, this FEM-augmented registration framework enhances geometric tracking and biomechanical evaluation accuracy, providing clinically relevant insights into patient-specific AV deformation to support individualized medical and intervention planning.

\end{abstract}

\begin{keyword}
Aortic valve biomechanics,
Finite element simulation,
Image registration,
Computational biomechanics

\end{keyword}


\maketitle

\section{Introduction}
The aortic valve (AV) plays an essential role in cardiovascular function by regulating blood flow from the left ventricle to the aorta and preventing regurgitation during diastole. In most healthy individuals, the trileaflet aortic valve (TAV) consists of three leaflets that undergo large, rapid, and highly anisotropic deformations throughout each cardiac cycle. These deformations expose valvular interstitial cells (VICs) to cyclic strains that regulate extracellular matrix (ECM) and maintain tissue homeostasis \citep{west2023effects, huang2007situ,yap2010dynamic,ayoub2017regulation}. Strain thus serves as the critical mechanobiological link connecting valve structure and the hemodynamic forces imposed on the valve to cellular behavior and long-term tissue remodeling. When strain patterns become abnormal (e.g., due to age-related degeneration, altered hemodynamics, or congenital defects), VICs activation and ECM dysregulation accelerate disease progression, as seen in calcific aortic valve disease (CAVD) \citep{davies1996demographic,yap2010dynamic,ayoub2017regulation,ayoub2021role,lam2016valve,decano2022disease,aggarwal2014architectural}.

Among congenital defects, bicuspid aortic valve (BAV) provides a clear example of how altered strain drives pathology. BAV, affecting 1–2\% of the population \citep{evangelista2011bicuspid}, results from fusion of adjacent leaflets, producing asymmetric geometry and disordered loading. These structural changes elevate localized stresses and strains, particularly near the leaflet raphe \citep{jermihov2011effect, merritt2014association, rego2022patient}. These mechanical abnormalities contribute to early ECM disorganization, rapid calcification, and accelerated onset of aortic stenosis (AS) or regurgitation (AR), often decades earlier than in normal tricuspid valves \citep{rutkovskiy2017valve,khang2023three}. Given that strain plays a central role in regulating cellular responses and long-term tissue remodeling, accurate patient-specific quantification of leaflet strain is critical. Specifically, strain can reveal disease progression and identify tissue regions at highest risk for dysfunction or failure. Although not yet part of routine clinical practice, patient-specific strain assessment could ultimately guide surgical decisions toward more durable repairs and reduced need for reintervention.

Despite this clear clinical need for precise and reliable strain assessment, in vivo strain estimation remains a significant unsolved challenge. Clinical imaging modalities, including 4D computed tomography (CT) and 4D echocardiography, lack sufficient temporal and spatial resolution to capture the extremely rapid geometric changes that occur during valve opening and closure \citep{meredith2025aortic}. Because conventional image-based registration algorithms fundamentally rely on incremental motion between sequential frames, the sparse sampling during fast deformation phases leads to large tracking errors. Additionally, limited soft tissue contrast and acoustic shadowing further degrade the accuracy of deformation estimates \citep{meredith2025aortic}. As a result, existing registration approaches struggle to propagate leaflet geometry across the full cardiac cycle and cannot reliably compute physiologically meaningful strain fields.

Since image-based tracking alone cannot recover full-cycle deformation, finite element (FE) and fluid–structure interaction (FSI) modeling offer a complementary path forward, though these approaches carry their own significant limitations. Prior computational studies have provided important insights into aortic valve leaflet mechanics and hemodynamics \citep{labrosse2015subject, chen2022image, lior2023semi, jermihov2011effect, rego2022patient, marom2012fluid, yin2024fluid}. Labrosse et al. used 3D+t TEE to construct subject-specific FE models of normal aortic valves and simulate valve function to estimate leaflet stresses \citep{labrosse2015subject}. Rego et al. proposed a patient-specific computational pipeline to quantify leaflet deformation in adult trileaflet and bicuspid aortic valves using real-time 3D echocardiography \citep{rego2022patient}. Their approach relied on manual segmentation of fully open and fully closed valve configurations and employed FE-based closure simulations to map the open-state geometry to the imaged closed state, with pressure loading and corrective forces iteratively optimized to achieve shape matching. More recent CT-based approaches have enabled semi-automated reconstruction of patient-specific volumetric aortic valve geometries from angiographic images \citep{lior2023semi}. In that work, volumetric leaflet models were constructed for a cohort of 25 pediatric patients; however, FSI simulations were demonstrated for only a single representative case and were primarily used to illustrate flow waveforms, without quantitative evaluation of leaflet deformation or patient-specific strain. Despite these advances, most FE and FSI frameworks rely on idealized or parametric geometries \citep{jermihov2011effect, yin2024fluid}, simplified boundary conditions \citep{marom2012fluid}, and high computational cost, and are often evaluated in small cohorts or single representative cases. Recent efforts have begun to explore tighter integration between biomechanics and image analysis. Wu et al. proposed a framework in which FE-simulated intermediate states were used to augment deep learning-based deformable image registration, demonstrating feasibility on 2D echocardiographic images of three pediatric tricuspid valves \citep{wu2024physics}. While promising, this initial example was limited to demonstrating shape matching across cardiac phases in two-dimensional images and did not incorporate strain quantification.

Forward FE and FSI simulations can estimate full-cycle valve deformation but cannot fully account for patient-specific variability in geometry, material properties, and boundary conditions. Image-based tracking, meanwhile, cannot capture the rapid motion of valve leaflets between frames. These complementary limitations suggest a natural integration: FE simulation can generate mechanically consistent intermediate deformation states missing from sparse imaging sequences, while image registration can correct deviations arising from modeling assumptions. Building on this idea, we introduce a patient-specific computational framework that couples finite element simulation with deformable image registration. The proposed approach relies only on manual segmentation of the open valve configuration to construct the FE model and uses this model to generate biomechanically informed intermediate configurations, facilitating robust alignment between anatomically distant structures at two time frames, such as fully open and fully closed valves. Image registration, in turn, corrects deviations arising from modeling assumptions and ensures that leaflet geometry accurately reflects patient-specific imaging data. By combining the strengths of mechanics-based modeling and image-based registration, this framework enables robust tracking of aortic valve deformation and accurate computation of leaflet strains, overcoming the fundamental limitations of imaging-only and simulation-only approaches. We further evaluate registration accuracy and compute strain distributions across multiple imaging modalities, including 4D CT and 4D TEE, and across a diverse cohort of adult trileaflet, adult bicuspid, and pediatric aortic valves. Ultimately, this work establishes a new pathway for precise patient-specific biomechanical assessment, enabling quantitative strain measurements that may inform tissue-level risk and disease progression.

\section{Methods}

\begin{figure*}[!h]
  \centering
  \resizebox{1 \textwidth}{!}{
   \includegraphics{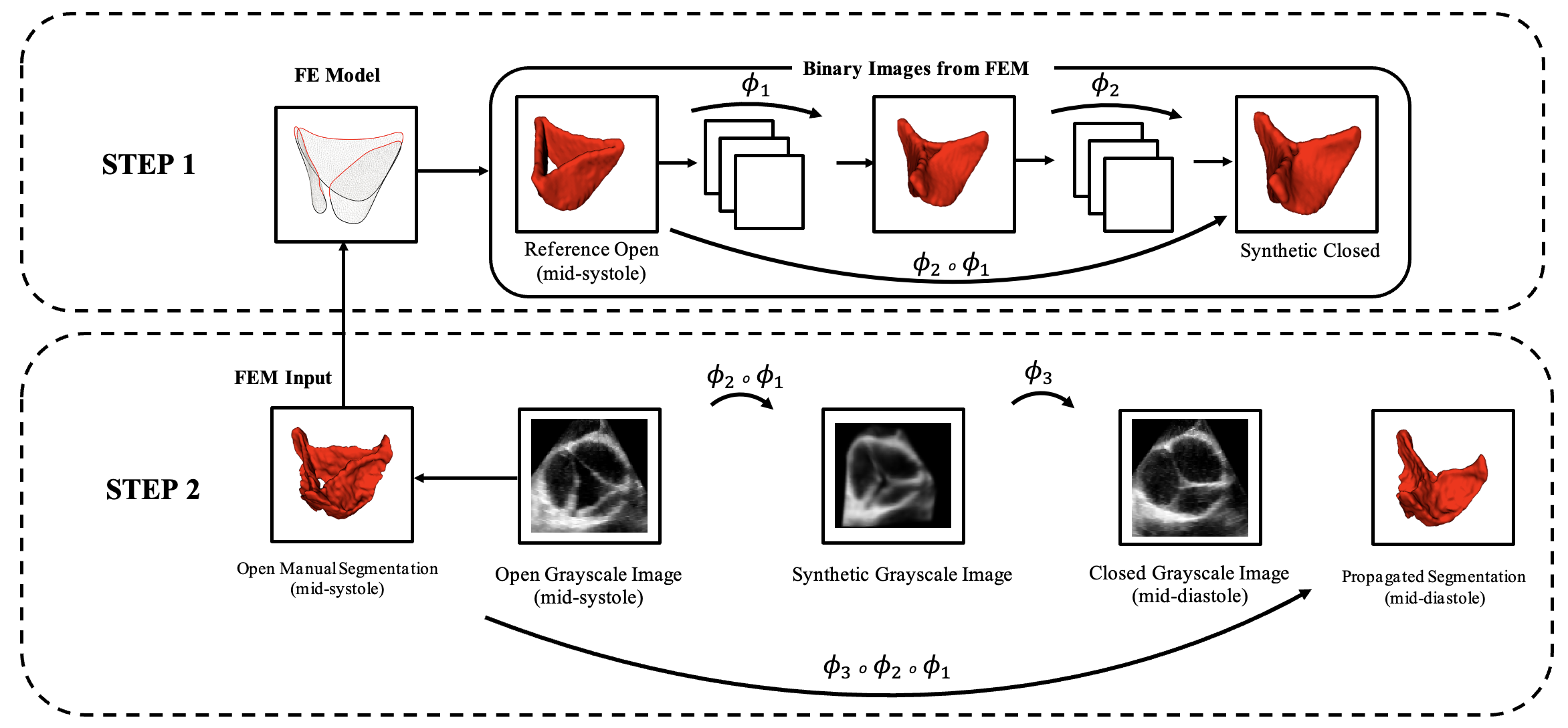}}
\caption{Registration pipeline with binary images from FE simulation.}\label{fig:registration-flowchart}
\end{figure*}

Conventional registration methods cannot accurately align the aortic valve between its two extreme configurations, mid-systolic (AV open) and mid-diastolic (AV closed), because the valve’s complex geometry undergoes rapid movement, substantial non-linear deformation, and significant shape changes throughout the cardiac cycle. As a result, conventional feature-tracking approaches have difficulty obtaining anatomically consistent models throughout the cardiac phases. To address this limitation, we proposed a hybrid framework that integrated biomechanical simulation using an FE model with image-based registration, as shown in Figure \ref{fig:registration-flowchart}.

The process begins with the manual segmentation of the aortic valve in the mid-systole frame (open reference configuration). In the first step, the medial surface of the valve leaflets is extracted from the mid-systole segmentation (AV open) and used to generate a finite element model to simulate valve closure. To characterize the biomechanical behavior of the valve tissue under physiological loading, material properties and diastolic pressure values were obtained from literature and the patient database (when available). Furthermore, physiologically informed boundary conditions obtained from direct registration and were applied to the leaflets to simulate the valve’s deformation from the mid-systole configuration (AV open) to the mid-diastole configuration (AV closed).
The simulation generates a sequence of intermediate geometries that approximate the valve’s movement between cardiac phases, from the mid-systolic configuration to a closed configuration near mid-diastole, which we refer to as the “synthetic closed” state. In the second step, these intermediate geometries serve as prior knowledge within the registration pipeline, guiding the propagation of mid-systolic segmentation across 4D TEE or  4D CT images. As a result, we obtain an image-driven model of the aortic valve and can evaluate the strain throughout the cardiac phases. In the following, we go through these steps in detail.

\subsection{Model preparation from clinical images}

\subsubsection{Segmentation}
An accurate geometric representation of the aortic valve is a crucial step in developing a physiologically relevant computational model for biomechanical simulation and deformation analysis. In this study, manual segmentation was used to extract the aortic valve anatomy from 4D TEE or 4D CT images acquired throughout the cardiac cycle. In total, we curated 14 4D TEE datasets and 6 4D CT datasets, 4 of which were from pediatric patients. The collection and analysis of these data were approved by the Institutional Review Boards at the Children’s Hospital of Philadelphia and the University of Pennsylvania. Two specific frames were selected for segmentation: (1) the mid-systolic frame, representing the fully open configuration of the valve, and (2) the mid-diastolic frame, representing the fully closed configuration. These frames capture the two extremes of valve deformation during cardiac cycle and provide crucial reference states for both simulation and registration.

Although ML-based segmentation methods have demonstrated good performance in many cardiac imaging applications, their accuracy depends heavily on image characteristics and on training data that reflect similar anatomical features and acquisition settings. To ensure the highest level of accuracy and consistency in the geometric models used for our analysis, we elected to perform manual segmentation for all patients.

Manual segmentation was performed using ITK-SNAP and 3D Slicer, two widely used software platforms for anatomical image labeling \citep{Lasso2022_SlicerHeart, py06nimg,fedorov20123d}. The open-frame segmentation served as input for extracting the leaflet surface and developing the FE model, while the closed-frame segmentation served as a reference or “ground truth” for validating propagated segmentations. Because the spatial resolution was limited by the original image acquisition, careful attention was required during segmentation to preserve anatomical details. To preserve key morphological features such as leaflet curvature, commissures, and attachment regions, each frame was manually traced slice by slice across multiple 2D planes, carefully reviewed in orthogonal views, and processed with minimal smoothing to keep small anatomical details accurate. Label maps were then interpolated into 3D volumes, with anatomical landmarks such as the valve annulus and sinus borders guiding the contours to maintain consistency across the two frames. Segmentation was performed under the supervision of a clinical expert to reduce operator bias and ensure accurate representation of the native leaflet anatomy for physiologically realistic simulations. The final open-frame label maps were converted into surface meshes for further modeling steps.

\subsubsection{Medial surface extraction}

After segmenting the valve in the mid-systolic (open) configuration, the medial surface was extracted from the segmentation to serve as the input geometry for FE analysis (Figure \ref{fig:FEM} (A)). The medial surface, representing the mid-surface of the valve leaflets, was generated using an interactive skeletonization approach applied to the surface mesh of the binary segmented volume, implemented in 3D Slicer \citep{herz2024open}. For completely fused leaflets, the two leaflets were modeled as a single entity to maintain geometric consistency in the medial surface representation.

The medial surface reduces the computational complexity of the FE model while preserving essential anatomical features such as leaflet curvature, commissures, the annular boundary, and the free edges. It also provides a mid-surface reference, from which thickness profiles can be defined across the leaflet to generate shell elements that accurately reflect the leaflet’s 3D geometry and mechanical behavior. 

Post-processing of the medial surface included smoothing with preserved boundaries, which were carried out using the FEBio software \citep{maas2012febio}. The result was a continuous, smooth 3D surface that accurately captured the valve geometry in the open reference configuration. This process ensures that the mesh topology is suitable for finite element simulation. This surface was later used to generate a shell-based finite element model, which serves as the initial configuration in simulating valve closure (Figure \ref{fig:FEM} (B)).

\subsection{Finite element valve model}

\subsubsection{Constitutive model}

The non-linear, nearly-incompressible mechanical behavior of aortic valve tissue was modeled using a hyperelastic constitutive model. Specifically, we employed the isotropic Lee--Sacks model, a Fung-type exponential strain-energy function previously developed for heart valve simulations \citep{sun2005finite, fan2014simulation}. This model combines a neo-Hookean matrix with a Fung-type exponential term to capture the nonlinear response of soft biological tissues under large deformations. In addition, it separates the contributions of isochoric deformations and volumetric deformations, allowing the energy associated with shape changes ($\Psi_\text{isochoric}$) and volume changes ($U_{vol}$) to be considered independently. The total strain-energy density can be expressed as,
\begin{align}
\Psi(C, J) &= \Psi_\text{isochoric}(c_0, c_1, c_2, I_1) + U_{vol}(J) 
\end{align}
where the isochoric strain-energy density defined as,
\begin{align}
 \Psi_\text{isochoric} = \frac{c_0}{2} (I_1 - 3) + \frac{c_1}{2} \left( e^{c_2 (I_1 - 3)^2} - 1 \right). 
\end{align}
$C$ is the right Cauchy-Green deformation tensor representing the full deformation, $I_1 = \mathrm{tr}(C)$ is its first invariant, and $J = \det(F)$ denotes the local volumetric change. The parameter $c_0$ represents the shear modulus of the neo-Hookean matrix, which governs the baseline linear elasticity of the tissue. The parameters $c_1$ and $c_2$ control the magnitude and rate of exponential stiffening, respectively, and are key to capturing the nonlinear behavior of collagenous soft tissues.

For these materials, the entire bulk (volumetric) behavior is determined by the function $U_{vol}(J)$. This function is constructed to have a value of 0 for $J = 1$ and to be positive for all other values of $J > 0$. In quadratic form, it can be written as $U_{vol}= \frac{k}{2} (J - 1)^2$, where $k$ is the bulk modulus, controlling the tissue’s resistance to volumetric changes and ensuring near-incompressibility, which is a common property of biological soft tissues.

The material parameters used in this study were $k = 5000$ kPa, $c_0 = 67$ kPa, $c_1 = 13$ kPa, and $c_2 = 35$. These values were chosen based on prior studies on material properties of heart valves \citep{wu2018anisotropic}, where parameters of the Lee–Sacks material model were determined by fitting equibiaxial experimental data reported in \citep{sun2005finite,lee2014inverse}. 

While the Fung-type model offers realistic mechanical behavior, it is known to introduce convergence challenges under certain loading conditions. Kamensky et al. \citep{kamensky2018contact} reported that the exponential term may cause these convergence difficulties, and therefore, they introduced a tangent scale factor which can improve numerical stability. In our simulations, a tangent scale factor of one was applied ($TF=1$), but model convergence was monitored carefully to mitigate any divergence issues.

Although recent studies such as \citep{armfield2024effect} have highlighted the importance of anisotropic behavior in valve leaflets due to collagen fiber alignment, we assumed isotropic material behavior for the sake of simplicity. It is acknowledged that this assumption introduces some degree of approximation, particularly in commissural regions where fiber orientation affects local deformation. However, in this study, the FE simulation is not intended to predict exact valve deformation but rather to generate intermediate valve configurations that serve as a structural prior for guiding the registration process. We assume that minor inaccuracies introduced by the isotropic assumption and uniform material properties for all the leaflets are compensated during the image-based registration step.

\subsubsection{Boundary conditions and loading}
\begin{figure}[!h]
\centering
\resizebox{1 \textwidth}{!}{%
\includegraphics{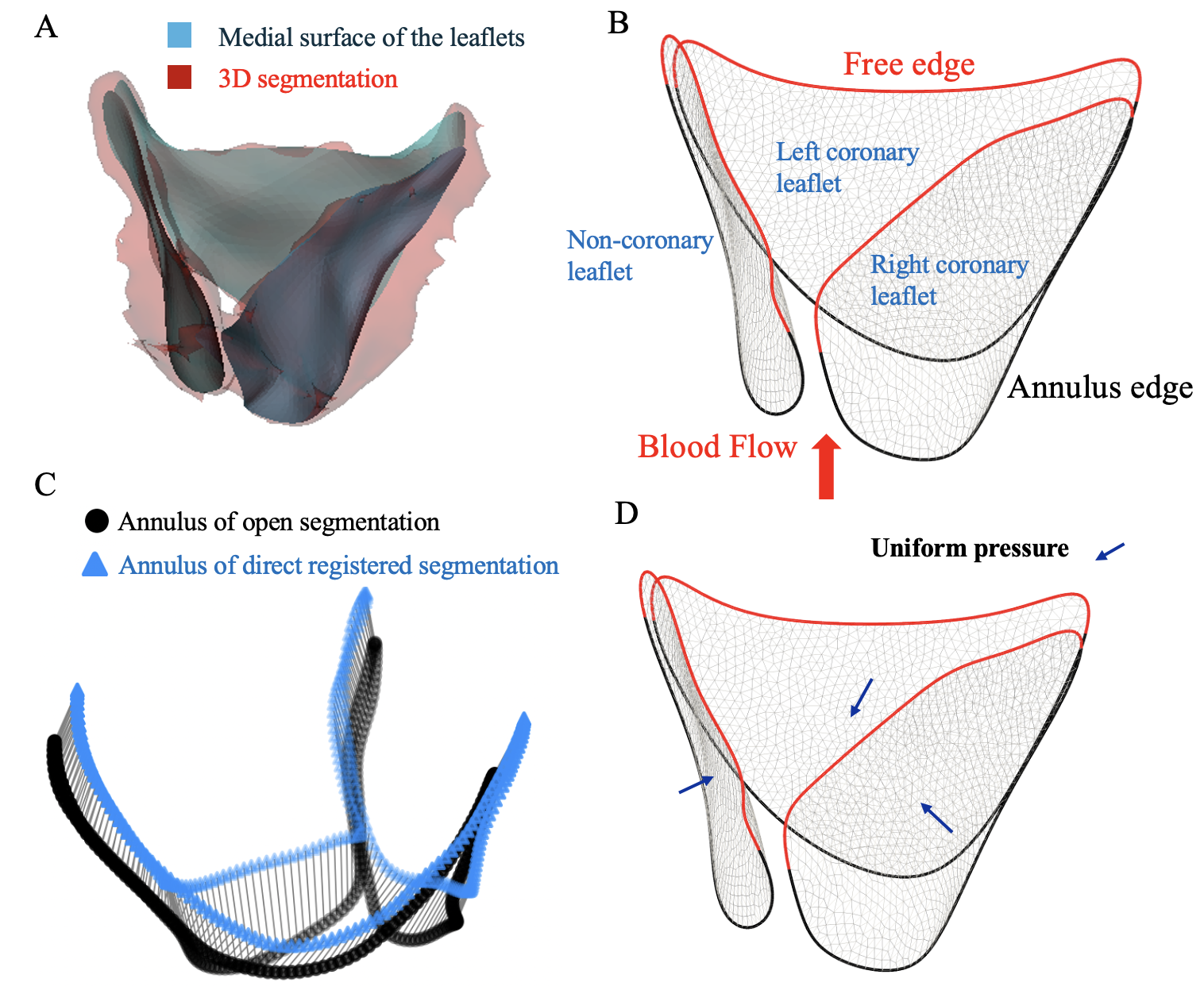}
}
\caption{Finite element model development. (A) Extraction of medial surface of the valve leaflets at mid-systole. (B) Definition of boundary conditions for the annulus and the free edge of the leaflet. (D) Estimation of annulus displacement. (C) Imposed uniform diastolic pressure on the leaflets.}
\label{fig:FEM}
\end{figure}

 Physiological boundary conditions (annulus movement and pressure loading on the valve leaflets) were applied to the reference open valve shell model to approximate the nearly closed configuration, referred to as the "synthetic closed" state. In particular, the annulus displacement during the cardiac phases was applied to the annulus to capture the dynamics of the aortic root. This displacement was obtained in a patient-specific manner from initial direct image registration. First, we directly registered the open mid-systolic grayscale image into the mid-diastolic image and we propagated the open segmentation in mid-systolic accordingly to obtain the segmentation in mid-diastolic frame. Therefore, we can approximate the annulus displacement between the two configurations (Figure \ref{fig:FEM} (C)) and impose it in our FE simulation. The free edges of the valve leaflets were left unconstrained.
 
Furthermore, pressure loading on the leaflets was applied to simulate diastolic closure. The transvalvular pressure gradient for AV, defined as the pressure difference between the left ventricle (LV) and the aorta (AO), (\(P=P_{LV}-P_{AO}\)), was reported over a cardiac cycle of 0.76 s  \citep{kim2008dynamic,hole1996hole}. They reported that during the opening phase ($t = 0$ s to $t = 0.18$ s), only a small pressure acted on the ventricular side of the leaflets, decreasing from about 4 mmHg to 0 mmHg. Once the closing phase began, pressure was applied on the aortic side and increased sharply between $t = 0.23$ s and $t = 0.3$ s, reaching approximately 80–100 mmHg at its peak. However, the actual pressure experienced by the leaflets during diastole is lower than systemic aortic diastolic pressure. Factors contributing to this include: (i)  the blood in the aortic sinuses moves in small circular patterns that create a soft fluid cushion (fluid cushioning), so the leaflets do not feel the full aortic pressure \citep{marom2012fluid}; (ii) leaflet geometry and coaptation mechanics, which create non-uniform load distribution; and (iii) patient-specific hemodynamics, notably in aortic regurgitation (AR), where retrograde flow during diastole reduces local pressure near the leaflets \citep{Yang2020_AR_DBP_RHR}. Clinical data show that adults with moderate-to-severe AR often have lower diastolic pressures, illustrating inter-patient variability \citep{Yang2020_AR_DBP_RHR}.

Simplified FE models that apply uniform pressure are known to overestimate leaflet stresses compared with fully coupled FSI simulations \citep{marom2012fluid}. Assuming a typical systemic diastolic pressure of approximately 80 mmHg in adults, the FSI study suggests that only about 90–95\% of this pressure is effectively transmitted to the leaflets due to fluid cushioning and coaptation dynamics \citep{marom2012fluid}. Accordingly, we selected a uniform aortic-side pressure of 75 mmHg ($\approx$ 94\%) to balance physiological realism, patient variability, and computational feasibility across all 16 adult cases (Figure~\ref{fig:FEM}D). For the pediatric patient, where diastolic blood pressure measured before CT was available, an average diastolic pressure of 45 mmHg was applied. It is important to note that the transvalvular pressure gradient reflects how hard the heart must work to push blood through the aortic valve and varies among patients depending on the severity of aortic stenosis or regurgitation \citep{khalili2022transvalvular}. Accordingly, the subsequent image registration step corrects small geometric deviations that may arise from applying a uniform pressure, ensuring that the resulting leaflet configuration remains consistent with the patient-specific imaging data.

\subsubsection{Simulation and model convergence}

The finite element model was solved using the dynamic implicit solver of FEBio \citep{maas2012febio}. The valve leaflets were modeled using linear triangular shell elements with a uniform thickness of 1.2 mm for adult patients and 1 mm for pediatric patients. The thickness value for aortic valve was obtained from \citep{sahasakul1988age}, where they reported mean thicknesses for aortic valve for different age groups. The simulations were performed multiple times on a single valve model using different mesh densities to assess convergence behavior. Mesh sizes were systematically reduced, and convergence in maximum displacement was achieved with a target edge length of 0.4 mm for the triangular elements, corresponding to a total of 8271 elements. The simulation generated a series of intermediate frames between the fully open and fully closed states, and approximated the "synthetic closed" configuration, which was exported to guide registration. To ensure reproducibility of the simulations, all model parameters are summarized in the Table \ref{tab:FEM_parameters}.

\begin{table*}[!h]
\centering
\caption{Finite element model parameters for adult and pediatric aortic valves}
\label{tab:FEM_parameters}

\resizebox{0.95\textwidth}{!}{%
\begin{tabular}{p{5cm} p{4.5cm} p{4.5cm}}
\hline
\textbf{Parameter} &
\begin{tabular}[c]{@{}c@{}}
\textbf{Adult valves}\\
\textbf{(n = 16)}
\end{tabular} &
\begin{tabular}[c]{@{}c@{}}
\textbf{Pediatric valves}\\
\textbf{(n = 4)}
\end{tabular} \\
\hline

\begin{tabular}[c]{@{}l@{}}
\textbf{Constitutive model}
\end{tabular} &
\multicolumn{2}{c}{\hspace{- 3cm}  Isotropic Lee--Sacks hyperelastic model} \\
\hline

\begin{tabular}[c]{@{}l@{}}
\textbf{Material parameters}
\end{tabular} &
\multicolumn{2}{c}{$c_0 = 67$ kPa, $c_1 = 13$ kPa, $c_2 = 35$, $k = 5000$ kPa, $TF = 1$} \\
\hline

\begin{tabular}[c]{@{}l@{}}
\textbf{Element type}
\end{tabular} &
\multicolumn{2}{c}{\hspace{- 3cm}  Linear triangular shell elements} \\
\hline

\begin{tabular}[c]{@{}l@{}}
\textbf{Leaflet thickness}
\end{tabular} &\hspace{0.75cm}
1.2 mm &
\hspace{0.75cm} 1.0 mm \\
\hline

\begin{tabular}[c]{@{}l@{}}
\textbf{Solver}
\end{tabular} &
\multicolumn{2}{c}{\hspace{- 3cm}  Dynamic implicit solver (FEBio)} \\
\hline

\begin{tabular}[c]{@{}l@{}}
\textbf{Annulus boundary}\\
\textbf{condition}
\end{tabular} &
\multicolumn{2}{c}{Patient-specific annulus displacement from image registration} \\
\hline

\begin{tabular}[c]{@{}l@{}}
\textbf{Uniform pressure}\\
\textbf{(aortic side)}
\end{tabular} &
\hspace{0.75cm} 75 mmHg &
\hspace{0.75cm} 45 mmHg \\
\hline

\end{tabular}%
}
\end{table*}

\subsection{Image registration }
Standard image-registration methods struggle with the large deformation of the aortic valve between systole and diastole, leading to misaligned propagated segmentations. To address this, we developed a biomechanically informed registration framework to improve robustness in tracking the aortic valve feature across cardiac phases. First, FE simulations generate a sequence of intermediate valve configurations as binary images; second, these intermediate frames are used to guide the registration (see Figure \ref{fig:registration-flowchart}). 

In the first step, given the open manual segmentation at mid-systole, we develop a medial surface of the valve \citep{herz2024open} and simulate valve closure. This simulation brings the aortic valve toward its mid-diastolic configuration, which we refer to as the “synthetic closed” state. From the FE simulation, we extract three frames: the initial, middle, and the final simulation frame, and convert these into binary images. These binary images are then resliced and resampled into the grayscale image space. Then, pairwise registration between the frames is performed to compute the transformations that map one configuration to the next. The composed transformation ($\phi_2 \circ \phi_1$) therefore maps the mid-systolic open segmentation to the synthetic closed configuration.

In the second step, this FE-derived transformation ($\phi_2 \circ \phi_1$) is applied to the mid-systolic grayscale image to obtain a synthetic closed image that closely approximates the actual mid-diastolic frame. This facilitates the final registration from the synthetic closed to the actual closed grayscale image in mid-diastole ($\phi_3$). The full composed transformation ($\phi_3 \circ \phi_2 \circ \phi_1$) is then used to propagate the manual open segmentation from mid-systole into the closed configuration at mid-diastole.

The registration was performed between the FE frames using affine and deformable registration with Greedy \citep{yushkevich2016ic} to obtain a composed transformation mapping the open configuration to the closed configuration. For affine registration, Greedy optimized an affine transformation matrix, initialized by matching the centers of the fixed and moving images (-ia-image‑centers), using a sum of squared differences (SSD) metric for segmentation and a normalized cross-correlation (NCC) metric for grayscale images. The NCC metric used a local kernel (radius 2 voxels), and the registration employed a multi-resolution schedule of 100×50×10 iterations across three resolution levels (coarse, intermediate and full resolution). For deformable registration, Greedy computed a dense displacement field (warp), initialized from the affine result (-it affine.mat), optimized using the same multi-resolution schedule, SSD or NCC depending on the image type, and a stationary velocity field model with Lie bracket correction (-svlb) which applies a more precise update to the velocity field, improving the accuracy of the resulting deformations. The computed warp and affine transformations were then used to resample both grayscale images and segmentation maps into the fixed frame, with label interpolation (-ri LABEL 0.2vox) applied to segmentations to minimize artifacts along boundaries.

This registration process ensures that the propagated segmentation remains aligned with patient anatomical images and eliminates the errors that could arise from simulation assumptions. The goal is to enable the construction of a precise, image-driven valve model throughout the cardiac phases.

\subsection{Strain calculation}
We compute the total strain on the leaflet for each subject by summing the strains from both steps. For the first step, we compute both the areal strain and the Green--Lagrange strain at the synthetic closed configuration obtained from the FE simulation. The areal strain for each triangular element in the FE mesh is defined as the relative change in its surface area between two configurations,
\begin{equation}
\varepsilon_A = \frac{A_{\text{d}} - A_{\text{r}}}{A_{\text{r}}},
\end{equation}
where $A_{\text{r}}$ is the area of the element in the reference configuration (open manual segmentation)  and $A_{\text{d}}$ is the area of the same element in the deformed configuration (synthetic closed). For the second step, we compute the deformation between the synthetic closed and the closed configuration. The total areal strain is obtained as the sum of the areal strain components from both steps,
\begin{equation}
\varepsilon_{A,\text{total}} = \varepsilon_{A,1} + \varepsilon_{A,2}.
\end{equation}

The Green-Lagrange strain tensor $\mathbf{E}$  is computed from the deformation gradient  as,
\begin{equation}
\mathbf{E}= 
\frac{1}{2}\left(\mathbf{F}^{\mathrm{T}}\mathbf{F} - \mathbf{I}\right),
\end{equation}
where $\mathbf{F}$ is the deformation gradient computed within the FE solver, and $\mathbf{I}$ is the identity tensor. However, to compute the Green–Lagrange strain in the second step, we calculate the in-plane Green–Lagrange strain directly from the nodal coordinates of the shell meshes in the two configurations. In the first step, the FE simulation deforms the leaflets from the open configuration to a near-closed state at mid-diastole, during which most of the deformation occurs, particularly out-of-plane bending. In the second step, the leaflets are already closed, and only small deformations, mainly in-plane, arise from the registration process, which corrects the leaflet geometry and aligns it with the patient-specific anatomy. Because the  deformations in the second step are small, it is reasonable to approximate the strain as in-plane, and this assumption is verified. Finally, the total Green–Lagrange strain is obtained by summing the FE-derived Green–Lagrange strain from the first step with the in-plane strain from the second step,
\[
\mathbf{E}_{\mathrm{total}} = \mathbf{E}_{\mathrm{FE}} + \mathbf{E}_{2D}^{(3D)},
\]
where, \[\mathbf{E}_{2D}^{(3D)} = 
\begin{bmatrix}
E_{xx} & E_{xy} & 0 \\
E_{xy} & E_{yy} & 0 \\
0 & 0 & 0
\end{bmatrix}.\]

The effective (von Mises--type) strain, a scalar measure of deformation, is computed from the total strain tensor $\mathbf{E}_{\mathrm{total}}$ as,
\[
\varepsilon_{\mathrm{eff}} = \sqrt{\mathbf{E}' : \mathbf{E}'},
\]
where, 
\[
\mathbf{E}' = \mathbf{E}_{\mathrm{total}} - \frac{1}{3} \mathrm{tr}(\mathbf{E}_{\mathrm{total}}) \mathbf{I},
\]
is the deviatoric part of the strain tensor.  In addition, the Green--Lagrange strain magnitude, which provides an overall measure of local deformation including volumetric changes, is defined as
\[
\varepsilon_{\mathrm{GL}} = \sqrt{\mathbf{E}_{\mathrm{total}} : \mathbf{E}_{\mathrm{total}}}.
\]
Together, these scalar measures allow quantification of both deviatoric (shape-changing) and total (overall) deformation in the aortic valve tissue.

Finally, strain distributions were visualized using colormaps overlaid on the medial surface mesh, enabling the identification of regions of elevated strain, particularly near the commissures and coaptation lines. In this study, the fully open AV configuration was assumed to be the reference configuration and considered as strain-free, and strains were computed between the two extreme deformation states of the valve, namely the open and fully closed configurations. While this assumption simplifies the analysis, it is well recognized that the AV leaflets in vivo are not free of strain in the open state. Previous studies have demonstrated the presence of residual and functional strains, including in-plane stretching as well as shear and bending strains induced by blood flow and leaflet curvature during systole \citep{aggarwal2016vivo}. However, adopting the open configuration as a reference state provides a practical framework for quantifying relative strain patterns associated with valve closure. These results provide insights into the biomechanical behavior of the valve, support the assessment of pathological conditions such as leaflet stiffening or prolapse, and could potentially guide strategies for valve repair surgery.

\section{Results}
We applied the methodology to 20 patients (16 adult and 4 pediatric), including 14 4D TEE and 6 4D CT datasets. These images were acquired intraoperatively or immediately prior to the procedure. Patient characteristics, including valve morphology and severity of aortic regurgitation (AR) and stenosis (AS), are summarized in Table \ref{tab:patient_characteristics}. Overall, the combined cohort represented a broad spectrum of valve morphologies and disease severities, encompassing patients with normal to severely regurgitant valves and none to moderate stenosis.

\begin{table}[!h]
\centering
\caption{Patient characteristics aortic valve }
\label{tab:patient_characteristics}

\resizebox{0.9\textwidth}{!}{%
\begin{tabular}{p{4cm} p{4.5cm} p{4.5cm} p{4.5cm}}
\hline
\textbf{Characteristic} &
\textbf{Adult tricuspid (n = 13)} &
\textbf{Adult bicuspid (n = 3)} &
\textbf{Pediatric (n = 4)} \\
\hline
\textbf{Morphology} &
Trileaflet &
Bicuspid (variable fusion patterns) &
Trileaflet (n = 3), Unicuspid (n = 1; partial fusion) \\
\hline
\textbf{Regurgitation severity} &
\begin{tabular}[t]{@{}l@{}}
None: 6\\
Mild: 2\\
Moderate: 2\\
Moderate-to-severe: 1\\
Severe: 2
\end{tabular} &
\begin{tabular}[t]{@{}l@{}}
None: 1\\
Mild: 1\\
Severe: 1
\end{tabular} &
\begin{tabular}[t]{@{}l@{}}
Moderate: 1\\
Moderate-to-severe: 1\\
Severe: 2
\end{tabular} \\
\hline
\textbf{Stenosis severity} &
None: 13 &
\begin{tabular}[t]{@{}l@{}}
None: 1\\
Mild: 1\\
Moderate-to-severe: 1 (calcification)
\end{tabular} &
\begin{tabular}[t]{@{}l@{}}
None: 3\\
Mild: 1 
\end{tabular} \\
\hline
\end{tabular}%
}
\end{table}

\begin{figure}[!h]
\centering
\resizebox{1 \textwidth}{!}{%

\includegraphics{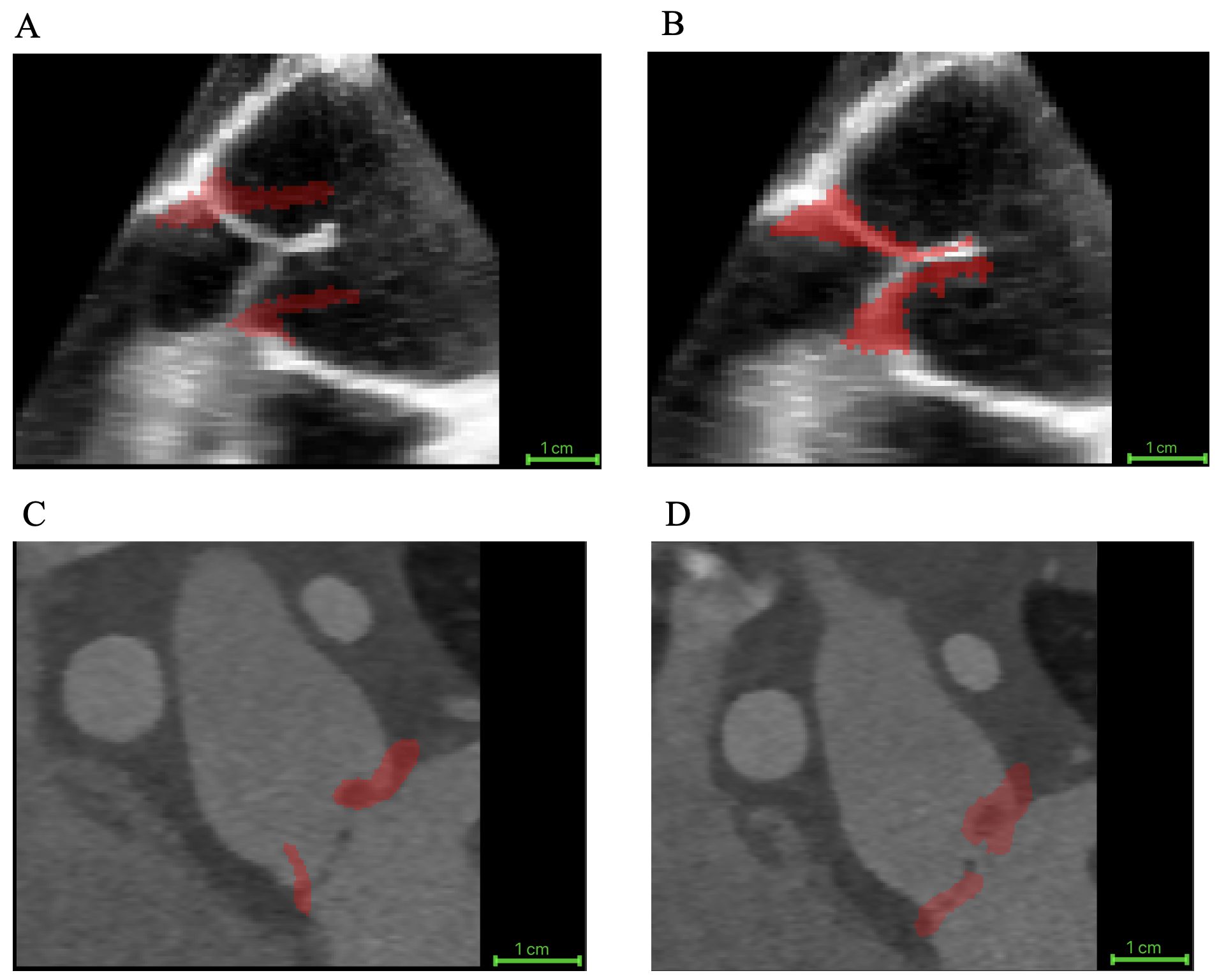}
}
\caption{Example of image-based registration results for one TEE case and one CT case. The reconstructed aortic valve leaflets at mid-diastole are shown in red. (A) TEE with direct registration, (B) TEE with FEM-augmented registration, (C) CT with direct registration, and (D) CT with FEM-augmented registration.}
\label{fig:registration-ct-TEE}
\end{figure}

The proposed biomechanically informed registration improved segmentation-tracking accuracy by an average of 40\%, reflecting the reduction in the mean distance between the tracked closed-state segmentation and the manual ground truth across all 20 cases when compared with direct registration. Specifically, this mean distance decreased from 3.70 ± 2.30 mm with direct registration (no FEM) to 2.23 ± 1.27 mm using our approach. Table \ref{tab:registration_summary} summarizes registration accuracy across valve morphologies and imaging modalities, comparing the mean distance between the manual ground truth and the propagated segmentation obtained with direct registration and proposed FEM-augmented registration. Figure \ref{fig:registration-ct-TEE} illustrates an example of the reconstructed patient-specific aortic valve closure (shown in red) overlaid on mid-diastole grayscale images. Panels A and B show a TEE case, and panels C and D show a CT case. For each modality, the left panel displays the result from direct registration, and the right panel shows the result from the FEM-augmented registration method. Furthermore, to assess the biomechanical behavior of different valve morphologies, we evaluated leaflet strain in adult trileaflet, adult bicuspid, and pediatric cases and compared the resulting strain ranges. The resulting strain maps exhibited physiologically consistent patterns.

\begin{table}[!h]
\centering
\caption{Registration accuracy across valve morphologies and imaging modalities. Values are mean distance between propagated closed-state segmentation and manual ground truth (mean ± SD).}
\label{tab:registration_summary}
\resizebox{0.95\textwidth}{!}{%
\begin{tabular}{l c c c c}
\hline
\textbf{Group} & \textbf{Modality} & \textbf{n} &
\textbf{Direct Registration (mm)} &
\textbf{Proposed Registration (mm)} \\
\hline
Adult trileaflet & 4D TEE & 11 & 2.52 ± 0.56 & 1.68 ± 0.52 \\
Adult bicuspid   & 4D TEE & 3  & 3.69 ± 0.34 & 2.04 ± 0.59 \\
Adult trileaflet & 4D CT & 2 & 7.47 ± 0.25 & 4.91 ± 0.83\\
Pediatric  & 4D CT & 4 & 6.44 ± 2.32 & 3.49 ± 1.68 \\
\hline
\textbf{All patients} & TEE \&  CT & 20 & 3.70 ± 2.30 & 2.23 ± 1.27 \\
\hline
\end{tabular}
}
\end{table}

\subsection{4D echocardiography images}

\subsubsection{Trileaflet aortic valves} 
\begin{figure}[!h]
\centering
\resizebox{0.8 \textwidth}{!}{%

\includegraphics{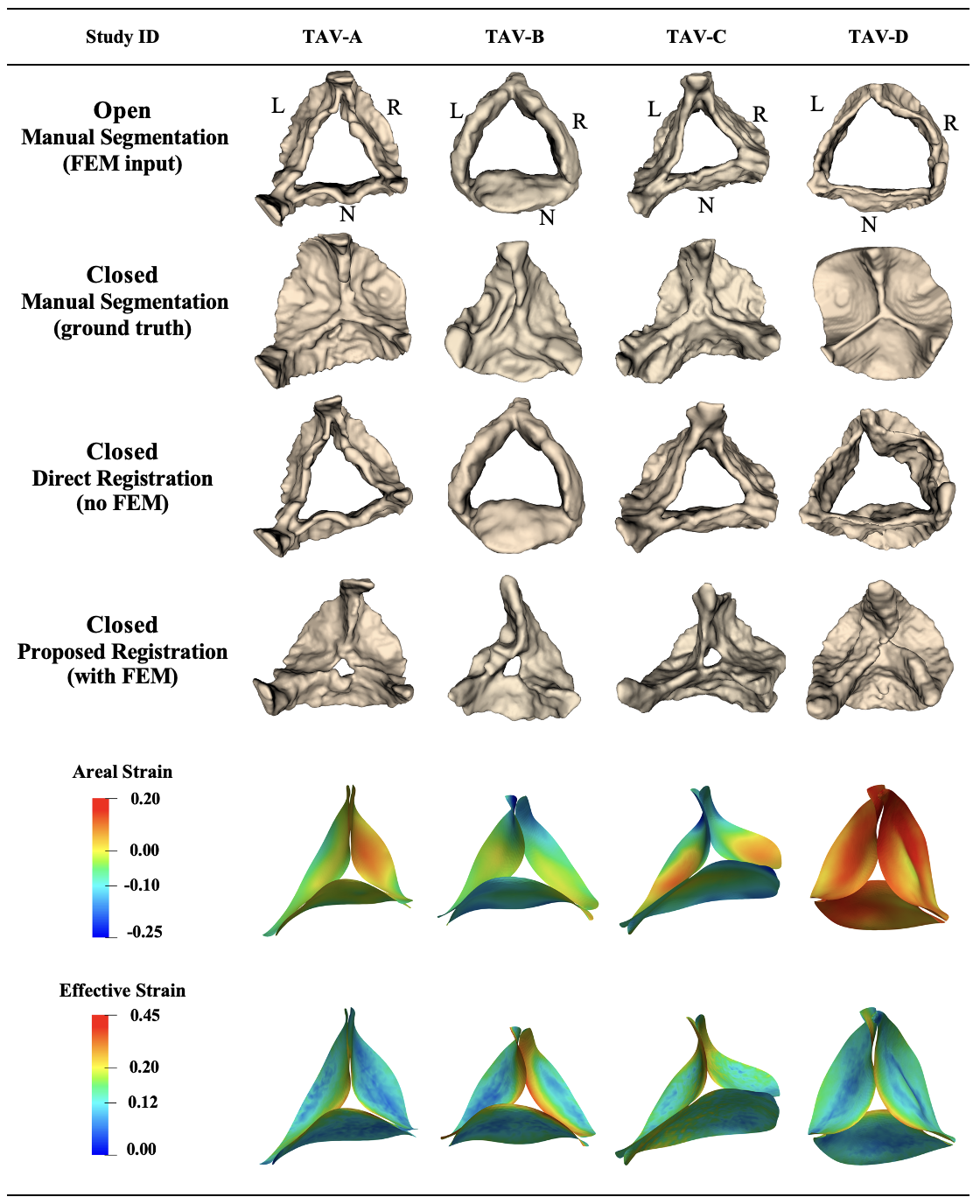}
}
\caption{Registration and strain results for trileaflet adult aortic valves from 4D TEE images. Examples with lower (TAV-A/B/C) and higher (TAV-D) mean distance values after valve registration from open to closed configuration, computed between registered and ground-truth manual segmentations. L: left coronary leaflet; R: right coronary leaflet; N: non-coronary leaflet.}
\label{fig:registration-tri}
\end{figure}

\textit{Patient characteristics:} Among 14 adult 4D TEE images, 11 were trileaflet and 3 were bicuspid aortic valves. The mean distance between the propagated segmentation for the closed state and the manual ground truth segmentation for these 11 patients was 1.68 ± 0.52 mm using the proposed biomechanically informed registration approach and 2.52 ± 0.56 mm with direct registration. Figure \ref{fig:registration-tri} shows the propagated segmentation results and strain for four of these cases. Patients TAV-A, TAV-B, and TAV-C were selected to represent some of the best registration results, whereas patient TAV-D was selected to represent one of the worst. These four adult trileaflet valves displayed a spectrum of valve function: normal (TAV-D), mild (TAV-C), mild-to-moderate (TAV-A), and moderate (TAV-B) aortic regurgitation, with no significant stenosis. Each case originated from a distinct clinical context: aortic aneurysm (TAV-A and TAV-C), aortic valve disorder (TAV-B), and coronary artery disease (TAV-D).

\textit{Registration accuracy:} The mean distance deviation from the ground truth for patients TAV-A, TAV-B, TAV-C, and TAV-D was 1.23, 1.27, 2.43, and 2.73 mm, respectively, using the proposed approach. For the same cases, the corresponding values using direct registration were 1.96, 2.08, 2.58, and 3.45 mm. Comparing the two methods, it was clear that direct registration (without FEM) could not fully reproduce leaflet closure. Among the 11 patients, one of the worst propagation results were observed for case TAV-D. Although the proposed method improved the mean distance by 21\% for this case, the large and complex deformation combined with limited image quality restricted further improvement in the final registration step. The large deformation in case TAV-D is evident from the comparison of fully open and fully closed segmentations, as well as the effective and areal strain values (12.2\% and 6.9\%, respectively), which are also visualized in Figure \ref{fig:registration-tri}.

\textit{Strain:} The mean effective Green-Lagrange strain across the leaflets for the representative adult trileaflet cases ranged from 11.7\% (TAV-A) to 14.4\% (TAV-B), with maximum local strain reaching 36.4\% in TAV-D. Maximum effective Green-Lagrange strain primarily occurred near coaptation lines and commissures, consistent with previous studies \citep{rego2022patient, abbasi2015leaflet, marom2012fluid, jermihov2011effect}. The mean areal strain ranged from -4.9\% (TAV-C) to 6.9\% (TAV-D), with maximum local areal strain reaching 26.5\% in TAV-D. Notably, TAV-D, despite a relatively large overall deformation, exhibited a uniform areal strain distribution consistent with normal valve function. In contrast, TAV-B, associated with mild-to-moderate regurgitation, had a lower mean areal strain (3.1\%) but a high maximum areal strain (22.1\%), reflecting that most of the leaflet surface underwent modest deformation while specific regions experienced localized overextension. Table \ref{tab:registration_strain_cases} summarizes the registration accuracy and strain values for these representative cases, providing a quantitative reference for the observations described above.

\begin{table}[!h]
\centering
\caption{Mean distance (MD) and strain values for representative cases shown in Figures 4--6. Strain values are averaged over all valve leaflets for each case. Maximum and minimum strain values observed on valve's leaflets are reported in parentheses.}
\label{tab:registration_strain_cases}
\resizebox{0.95\textwidth}{!}{%
\begin{tabular}{l c c c c c}
\hline
\textbf{Case} & \textbf{Valve type / Modality} &
\textbf{Direct MD (mm)} &
\textbf{Proposed MD (mm)} &
\begin{tabular}[c]{@{}c@{}}
\textbf{Mean effective strain (\%)}\\
\textbf{(min \& max)}
\end{tabular} &
\begin{tabular}[c]{@{}c@{}}
\textbf{Mean areal strain (\%)}\\
\textbf{(min \& max)}
\end{tabular}\\
\hline
TAV-A & Adult trileaflet / 4D TEE & 1.96 & 1.23 & 11.7 (2.15 \& 30.9) & -3.6 (-12.2 \& 8.0) \\
TAV-B & Adult trileaflet / 4D TEE & 2.08 & 1.27 & 14.4 (2.91 \& 34.2)  & 3.1 (-13.4 \& 22.1) \\
TAV-C & Adult trileaflet / 4D TEE & 2.58 & 2.43 & 14.2 (4.6 \& 28.1) & -4.9 (-19.8 \& 9.2) \\
TAV-D & Adult trileaflet / 4D TEE & 3.45 & 2.73 & 12.2 (2.9 \& 36.4) & 6.9 (-12.8 \& 26.5) \\
\hline
BAV-A & Adult bicuspid / 4D TEE & 3.38 & 1.62 & 13.2 (4.5 \& 41.9) & -3.2 (-26.7 \& 13.0) \\
BAV-B & Adult bicuspid / 4D TEE & 3.63 & 1.78 & 14.1 (3.5 \& 40.3) & -10.7 (-30.4 \& 17.6) \\
BAV-C & Adult bicuspid / 4D TEE & 4.06 & 2.71 & 13.0 (2.6 \& 23.4) & -14.0 (-26.4 \& -1.9) \\
\hline
A & Pediatric trileaflet / 4D CT & 2.97 & 1.91 & 15.5 (4.2 \& 34.3) & -1.8 (-24.0 \& 15.0) \\
B & Pediatric trileaflet / 4D CT & 4.50 & 2.30  & 15.7 (3.9 \& 37.8) & -0.6 (-25.5 \& 17.6) \\
C & Pediatric unicuspid / 4D CT & 9.62 & 5.12 & 15.8 (4.5 \& 45.0) & 2.4 (-31.1 \& 23.3) \\
\hline
\end{tabular}%
}
\end{table}

\subsubsection{Bicuspid aortic valves} 
\begin{figure}[!h]
\centering
\resizebox{0.8 \textwidth}{!}{%

\includegraphics{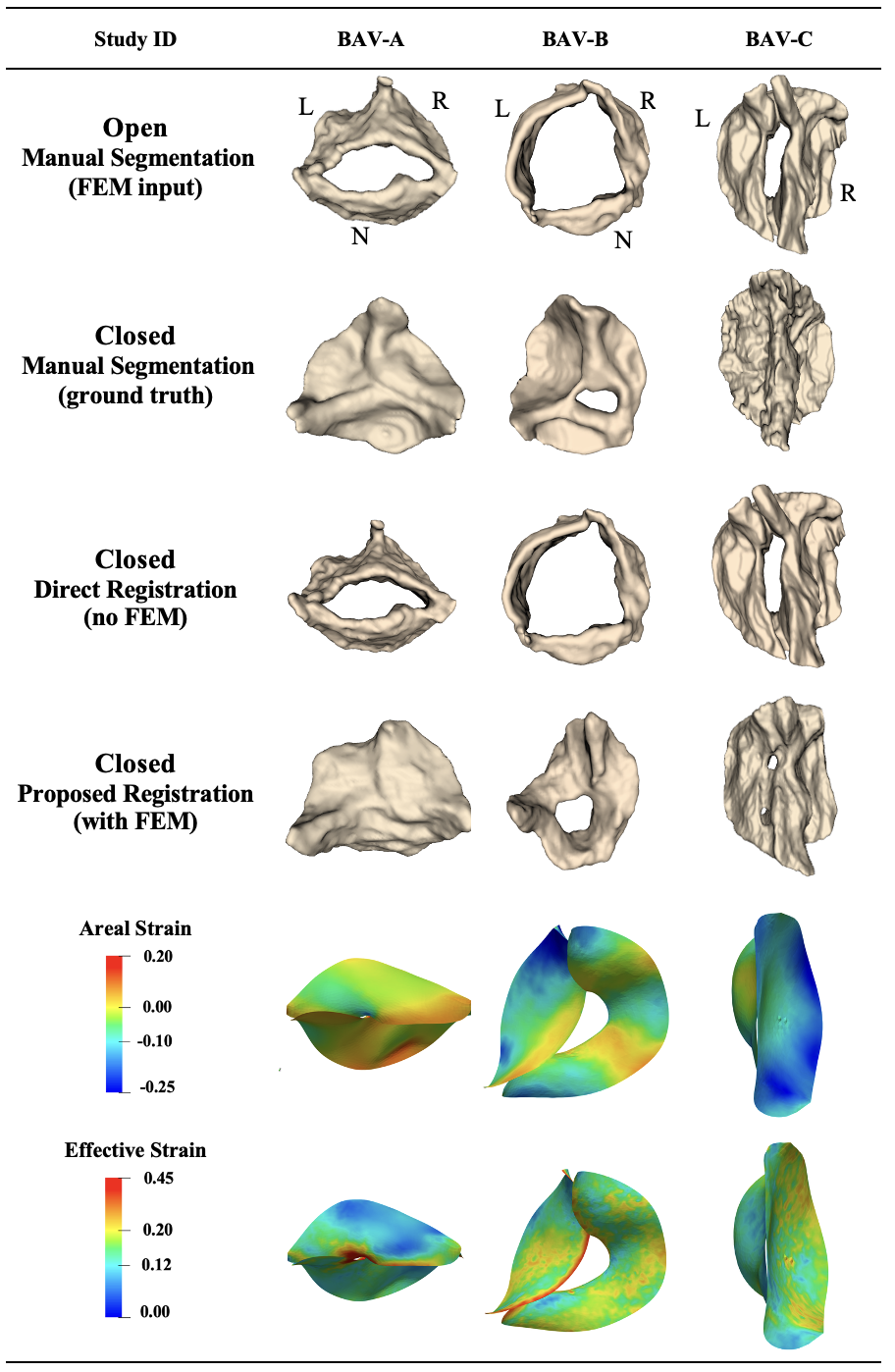}
}
\caption{Registration and strain results on bicuspid adult aortic valves 4D TEE images, showing valve registration from open to closed configuration. L: left coronary leaflet; R: right coronary leaflet; N: non-coronary leaflet. }
\label{fig:registration-bi}
\end{figure}

\textit{Patient characteristics:} Among 14  adult 4D TEE images, 3 were bicuspid aortic valves. The mean distance between the propagated segmentation for the closed state and the manual ground truth segmentation for these 3 patients was 2.04 ± 0.59 mm using the proposed biomechanically informed registration approach and 3.69 ± 0.34 mm with direct registration. Figure \ref{fig:registration-bi} shows the propagated segmentation results and strain for these cases. These three adult bicuspid valves exhibited a range of valve function: no aortic regurgitation (BAV-C), mild regurgitation (BAV-A), and severe aortic regurgitation (BAV-B). Aortic stenosis severity varied: absent in BAV-B, mild in BAV-A, and moderate-to-severe in BAV-C. Calcification was present in BAV-C (severe). Each case arose from a distinct clinical context: ascending aortic aneurysm (BAV-A), aortic regurgitation (BAV-B), and combined aortic aneurysm and aortic valve disease (BAV-C).

\textit{Registration accuracy:} The mean distance deviation from the ground truth for patients BAV-A, BAV-B, and BAV-C was 1.62, 1.78,  and 2.71 mm, respectively, using the proposed approach, and these metrics for the same cases were 3.38, 3.63, and 4.06 mm using direct registration. Patient BAV-B had severe regurgitation, which can be observed in the propagated segmentation using proposed method in Figure \ref{fig:registration-bi}.

\textit{Strain:} Furthermore, the mean effective Green-Lagrange strain across the leaflets for the bicuspid cases ranged from 13.0\% (BAV-C) to 14.1\% (BAV-B), with maximum local strain reaching 41.9\% in BAV-A, reflecting regions of high leaflet deformation during closure (Table \ref{tab:registration_strain_cases}). The mean areal strain varied more widely, from -14.0\% (BAV-C) to -3.2\% (BAV-A), with BAV-B showing an intermediate mean areal strain of -10.7\% and a high maximum of 17.6\%, indicating that most of the leaflet surface underwent moderate compression, while specific regions experienced local overextension. These patterns correlate with valve pathology: BAV-A, with a larger coaptation area, exhibited higher maximum effective strain; BAV-B, with severe regurgitation, showed pronounced local overextension; and BAV-C, with severe calcification, displayed lower maximum strain and uniform compression, consistent with restricted leaflet mobility  (Figure \ref{fig:registration-bi}).

\subsection{4D computed tomography images} 
\begin{figure}[!h]
\centering
\resizebox{0.8 \textwidth}{!}{%

\includegraphics{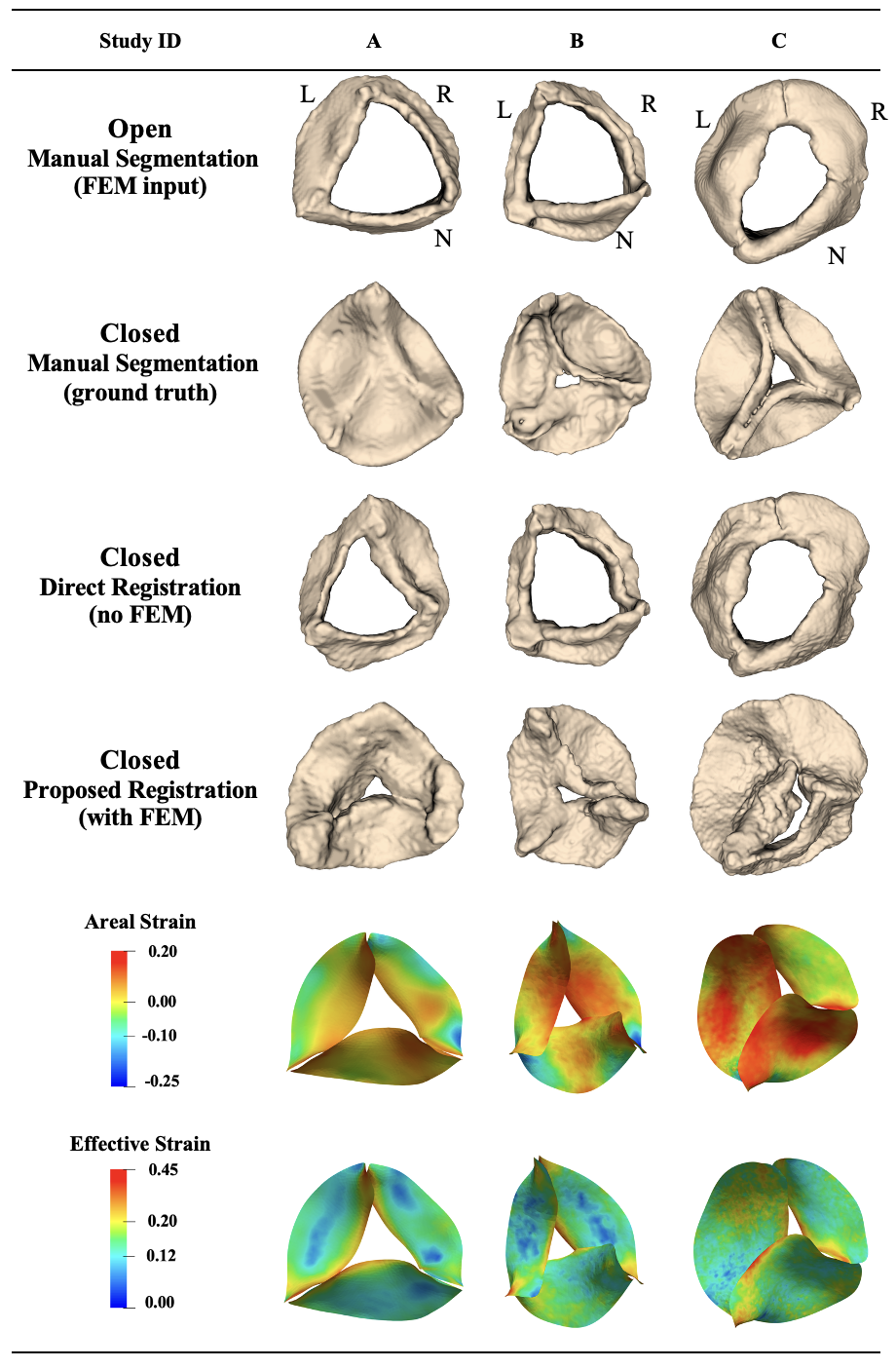}
}
\caption{Registration and strain results for pediatric aortic valves from 4D CT images, showing valve registration from open to closed configuration. L: left coronary leaflet; R: right coronary leaflet; N: non-coronary leaflet.}
\label{fig:registration-ct}
\end{figure}

\textit{Patient characteristics:} The 6 patients with 4D CT images included 4 pediatric patients and 2 adults. The mean distance between the propagated segmentation for the closed state and the manual ground-truth segmentation for these 6 patients was 3.49 ± 1.68 mm using the proposed biomechanically informed registration approach and 6.44 ± 2.32 mm with direct registration. Figure \ref{fig:registration-ct} shows the propagated segmentation results and strain for 3 pediatric cases. Patients A and B were selected to represent some of the best registration results, whereas patient C was selected to represent one of the worst. These three pediatric aortic valves exhibited a range of  function: trileaflet valves (A and B) and a unicuspid valve with partial R/N and R/L leaflet fusions (C). Aortic regurgitation severity ranged from severe (A) to moderate-to-severe (B) and moderate (C). Aortic stenosis was absent in A and B cases and mild in C. Each case arose from a distinct clinical context: severe neo-aortic insufficiency (A), moderate-to-severe neo-aortic regurgitation with moderate left ventricular dilation (B), and combined aortic valve insufficiency and stenosis (C).

\textit{Registration accuracy:} The mean distance deviation from the ground truth for patients A, B, and C using the proposed approach was 1.91, 2.30, and 5.12 mm, respectively. Using direct registration, the corresponding values were 2.97, 4.50, and 9.62 mm. The worst propagation performance was observed for case C (unicuspid). In this case, the mean distance remained high for both methods, indicating inferior image quality. Although the proposed approach improved the mean distance by 47\%, the large and complex leaflet deformations limited the accuracy of the FE simulation. Consequently, due to the limited image quality the final registration step also was unable to fully capture the valve motion.

\textit{Strain:} The mean effective Green–Lagrange strain across the pediatric valve leaflets ranged from 15.5\% (A) to 15.8\% (C), with maximum local strain reaching 45.0\% in the unicuspid case C, reflecting extensive leaflet deformation during closure (Table \ref{tab:registration_strain_cases}). The mean areal strain varied from -1.8\% (A) to 2.4\% (C), with case C showing the most heterogeneous deformation, including extreme local compression (-31.1\%) and high local expansion (23.3\%). In contrast, cases A and B (trileaflet) exhibited moderate areal strain values, with large local minima (-24\% to -25.5\%) and moderate maxima (15–17\%), consistent more uniform leaflet motion. These patterns illustrate that pediatric valves, particularly case C (unicuspid), undergo larger and more nonuniform deformations than adult valves.

\subsection{Strain evaluation on the leaflets}
\begin{figure}[!h]
\centering
\resizebox{1\textwidth}{!}{\includegraphics{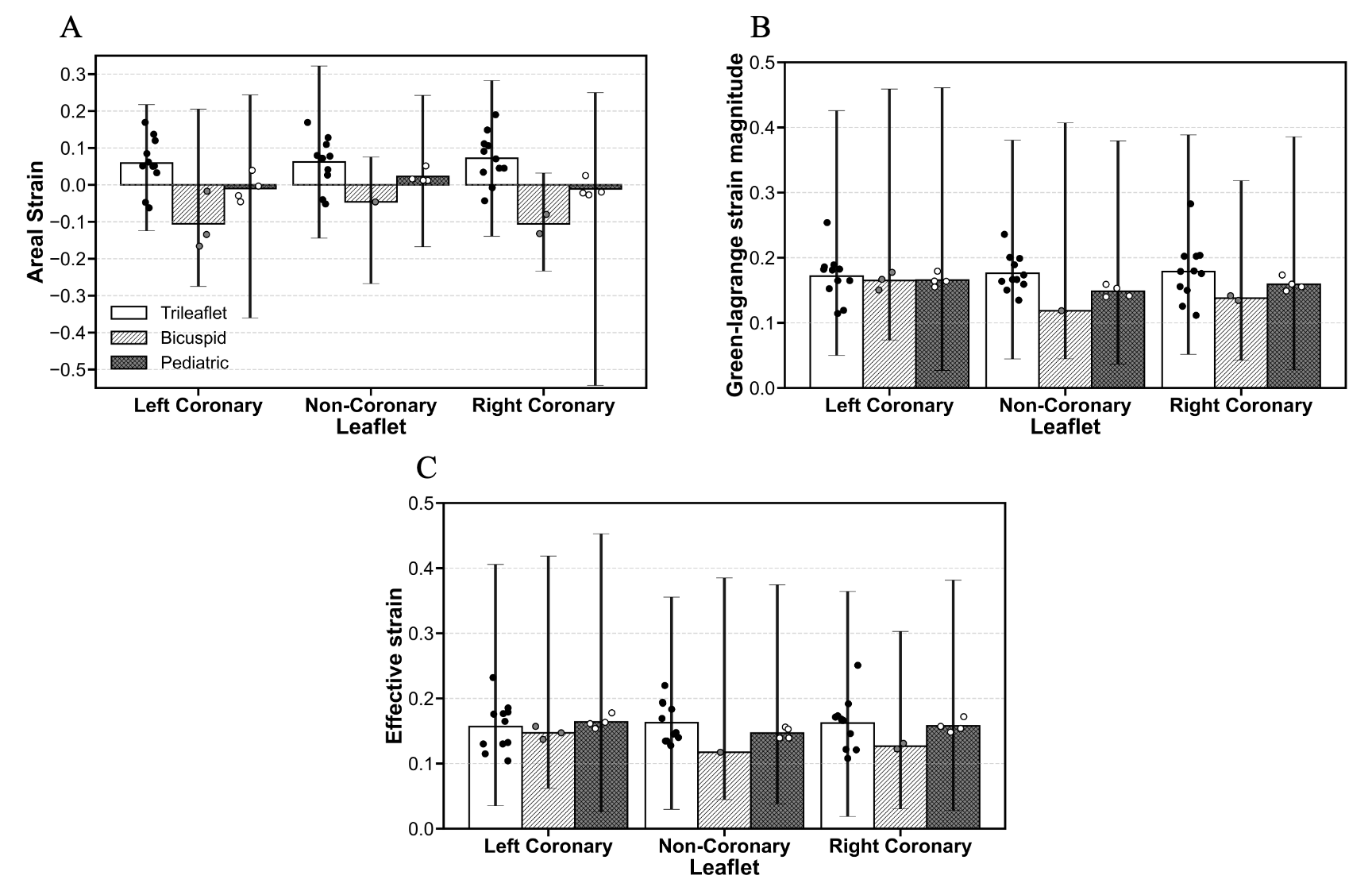}}

\caption{Strain analysis across the aortic valve leaflets (left coronary, non-coronary, and right coronary) in trileaflet, bicuspid, and pediatric valve groups. Bars show the mean strain for each leaflet within each valve group. Vertical ranges indicate the observed minimum–maximum strain across samples for each leaflet within each group. Individual points over each bar represent the mean strain measured on each valve leaflet. (A) Areal strain; (B) Green-Lagrange magnitude; (C) Effective strain. }
\label{fig:strain}
\end{figure}

Furthermore, we analyzed areal strain, the magnitude of the Green–Lagrange strain, and effective strain for each aortic valve leaflet (left coronary, non-coronary, and right coronary) in trileaflet adult, bicuspid adult, and pediatric valve groups, as shown in Figure \ref{fig:strain}. To facilitate comparison across valve groups and strain metrics, the bars in Figure \ref{fig:strain} represent the mean strain for each leaflet within each valve group, while the vertical lines indicate the observed minimum and maximum strain for each leaflet within each group. Individual points over each bar represent the mean strain measured on each valve leaflet.

The results for areal strain across aortic valve leaflets (left coronary, non-coronary, and right coronary) in trileaflet adult, bicuspid adult, and pediatric valve groups are shown in Figure \ref{fig:strain} A. In trileaflet adult valves, mean areal strain was 5.9\%, 6.2\%, and 7.2\% for the left coronary, non-coronary, and right coronary leaflets, respectively. The maximum strain observed in this group was 32.2\% and the minimum strain was -14.4\%, both on the non-coronary leaflet. In contrast, bicuspid adult valves exhibited lower, predominantly negative areal strains: -10.6\% (left coronary), -4.6\% (non-coronary), and -10.6\% (right coronary). The observed strain range was wide, from -27.5\% to 20.5\% on the left coronary leaflet, reflecting pronounced heterogeneity. Pediatric valves displayed generally lower areal strains compared with adult groups, with mean values near zero: -1.0\% (left coronary), 2.3\% (non-coronary), and -1.1\% (right coronary). Despite these low mean values, observed strain ranges were wide, particularly for the right coronary leaflet (-54.3\% to 25.0\%).

The results for Green-Lagrange magnitude strain, which captures the overall local deformation including both deviatoric and volumetric components, across the same leaflets and valve groups are shown in Figure \ref{fig:strain} B. In trileaflet adult valves, mean strains were 17.3\%, 17.7\%, and 17.6\% for the left coronary, non-coronary, and right coronary leaflets, respectively. The maximum observed strain was 42.6\% on the left coronary leaflet, and the minimum observed strain was 4.5\% on the non-coronary leaflet. Bicuspid adult valves exhibited slightly lower mean strains: 16.5\% (left coronary), 11.9\% (non-coronary), and 13.8\% (right coronary), with a maximum strain of 45.9\% on the left coronary leaflet and a minimum of 4.3\% on the right coronary. Pediatric valves showed mean strains comparable to adult groups: 16.6\% (left coronary), 14.8\% (non-coronary), and 15.9\% (right coronary). Similar to areal strain, Green-Lagrange strain magnitude exhibited wide ranges, particularly for the left coronary leaflet (2.7\% to 46.1\%).

Finally, we present the results for effective (von Mises–type) strain, which quantifies primarily deviatoric (shape-changing) deformation, across the same leaflets and valve groups (Figure \ref{fig:strain}C). Effective strain values followed patterns similar to those observed for Green–Lagrange magnitude strain but were consistently lower, reflecting the exclusion of volumetric contributions to the total deformation. Importantly, the difference between trileaflet adult and pediatric mean strain was reduced when assessed using effective strain. Effective strain values were similar between the two groups, with trileaflet adult leaflet averages ranging from 15.9 to 16.5\% and pediatric valve averages ranging from 14.7 to 16.4\%, resulting in minimal group differences. This finding suggests that age and size related differences in leaflet deformation are more associated with volumetric deformation than with deviatoric deformation.

\section{Discussion}
Reliable and precise estimation of patient-specific aortic valve leaflet strain is crucial to understand the link between valve biomechanics, disease progression, and tissue remodeling, since abnormal strain patterns accelerate extracellular matrix dysregulation and calcific aortic valve disease, particularly in bicuspid and pediatric valves with congenital defects \cite{jermihov2011effect,merritt2014association,rego2022patient,ayoub2017regulation,ayoub2021role}. Despite this clear clinical and mechanobiological importance, in vivo strain estimation remains a significant challenge. Advanced imaging and biomechanical modeling have independently improved our understanding of valve function, but neither approach alone can capture physiologically reliable and precise leaflet deformation. Image-based tracking methods are limited by sparse temporal sampling, low soft tissue contrast, and rapid leaflet motion, which lead to large registration errors \cite{meredith2025aortic}. FE and FSI models are constrained by idealized geometries, assumed boundary conditions and material properties, high computational cost, and limited patient-specific validation \cite{marom2012fluid,labrosse2015subject,rego2022patient,yin2024fluid}. To address this,  we developed a framework that couples deformable image-based registration with finite element simulation to accurately reconstruct patient-specific aortic valve closure and quantify leaflet strain. Unlike existing shape-matching or FE-based closure frameworks that rely on segmentation of both fully open and fully closed valve configurations and iteratively optimize pressure loading and corrective forces to achieve geometric correspondence, the proposed approach requires segmentation of only the open valve configuration. The FE model is used to generate biomechanically informed intermediate valve configurations that facilitate robust alignment between anatomically distant fully open and fully closed valve image frames through deformable image registration, thereby eliminating the need for direct shape matching to the closed segmentation. This integrated framework not only improves registration accuracy but also addresses uncertainties in boundary conditions and material properties during biomechanical analysis, while reducing reliance on fully detailed valve models that require many assumed parameters and incur high computational cost. The results of this study demonstrate that integrating these complementary approaches can substantially improve image tracking and enable more accurate estimation of leaflet deformation. By embedding finite element-derived intermediate configurations into deformable image registration, the proposed framework achieves robust tracking between fully open and fully closed valve states and provides consistent, patient-specific strain estimation across diverse valve groups, offering a practical pathway toward precise biomechanical characterization of aortic valve function.

\subsection{Image registration improvement}
Across 20 patients, we observed a 40\% improvement in registration accuracy using the proposed FEM-augmented tracking method compared to direct registration. This improvement reflects the reduction in the mean distance between the tracked closed-state segmentation and the manual ground truth across all 20 cases, calculated as the relative decrease compared to direct registration. When analyzed separately, the improvement was 33\% for the 14 4D TEE images and 46\% for the 6 4D CT images, calculated in the same way for each imaging modality (Table \ref{tab:registration_summary} ). The registration improvement reflects the benefit of incorporating intermediate frames from FE simulation, which were used as binary images to guide registration, whereas direct registration relies solely on image information.

\subsection{Physiologically meaningful areal strain patterns}
Areal strain analysis revealed distinct and physiologically meaningful patterns of leaflet deformation across patient groups. Trileaflet adult valves showed positive mean areal expansion across all leaflets, consistent with symmetric valve geometry and normal physiological loading. However, individual leaflet measurements included negative values, indicating contraction even in trileaflet valves. In contrast, bicuspid adult valves exhibited overall leaflet contraction with notable asymmetry: the left and right coronary leaflets showed comparable negative strain, while the non-coronary leaflet displayed smaller contraction. Despite local leaflet expansion up to 20.5\%, the overall pattern indicates restricted and uneven leaflet expansion, highlighting the mechanical consequences of fused leaflet geometry and supporting prior reports of asymmetric mechanical loading in bicuspid valves \citep{jermihov2011effect, merritt2014association, rego2022patient}. However, inter-leaflet variability in the bicuspid group should be made cautiously due to the limited number of available leaflets, particularly for the non-coronary leaflet. Pediatric valves showed near-zero mean areal strain but notable strain variability, particularly in the right coronary leaflet, likely reflecting ongoing growth and evolving valve geometry. The locations of extreme strain values across groups were consistent with expected deformation patterns as reported previously, with higher strain near the free edge and commissures and lower strain near the annulus \citep{aggarwal2016vivo}. 

\subsection{Green–Lagrange strain highlights local deformation differences}
Green–Lagrange strain magnitude analysis provided a full measure of local leaflet deformation and complemented the areal strain findings by highlighting differences in deformation across valve groups. Adult trileaflet valves exhibited a relatively uniform mean Green–Lagrange strain magnitude across all three leaflets, suggesting a balanced mechanical loading and relatively symmetric deformation under physiological conditions. Although local variability was present, strain magnitudes were broadly comparable among leaflets, consistent with structurally normal valve. Adult Bicuspid  valves showed reduced mean Green–Lagrange strain magnitudes, notably on the non-coronary leaflet, reflecting the asymmetric geometry and altered load sharing associated with leaflet fusion. Although the average strains were lower, bicuspid valves exhibited some of the largest observed maximum strain magnitudes, exceeding 45\% on the left coronary leaflet. This combination of reduced mean strain and elevated local maxima highlights the heterogeneous deformation in bicuspid valves, where localized regions may experience substantially higher deformation despite lower overall leaflet averages. This potentially challenges tissue hemostasis and contributes to rapid disease progression. Pediatric valves demonstrated mean Green–Lagrange strain magnitudes comparable to those of trileaflet adult valves, but with greater leaflet-to-leaflet variability and broader strain ranges. In particular, the left coronary leaflet in pediatric valves exhibited strains ranging from 3\% to 46\%, suggesting significant strain heterogeneity. This increased variability likely reflects developmental differences in tissue structure and leaflet geometry, which may contribute to less uniform mechanical behavior during valve function. The highest Green-Lagrange strain magnitudes were observed along leaflet free edges and near commissure regions, consistent with previous studies reporting strain magnitudes from 17\% to 54\% \citep{rego2022patient, abbasi2015leaflet, marom2012fluid, jermihov2011effect}. Overall, these findings confirm that trileaflet adult valves exhibit relatively uniform deformation, bicuspid valves show slightly reduced and asymmetric strain patterns, and pediatric valves display more heterogeneous behavior, underscoring the importance of patient-specific mechanical assessment, particularly for valves with asymmetric or developing geometry.

\subsection{Shear-dominated deformation} 
Effective strain analysis showed the deviatoric (shape-changing) component of leaflet deformation, reflecting changes in shape rather than total deformation. The effective (von Mises–type) strain is computed from the deviatoric part of the total strain tensor, which excludes volumetric contributions, while the Green–Lagrange strain magnitude measures the total local deformation, including volumetric changes. Across all patients and leaflet groups, effective strain values were only slightly lower than the corresponding Green–Lagrange strain magnitudes, indicating that the majority of total deformation is captured by the deviatoric (shear) component rather than by volumetric changes. Mean effective strains were more closely aligned between trileaflet adult and pediatric valves, and the differences between groups were smaller than those observed in the Green–Lagrange strain magnitudes. This suggests that size- and age-related differences in total deformation are primarily due to volumetric components. In structurally normal trileaflet valves, this indicates that the fundamental shear-dominated deformation mechanisms remain consistent across age groups, despite differences in valve size.

\subsection{Limitations}
Despite these promising results, several limitations should be acknowledged. First, the sample size was relatively small, with only 20 patients, which limits the generalizability of our findings and underscores the need for validation in larger cohorts. Second, while the proposed method improves registration accuracy, its performance is influenced by the quality and resolution of the input imaging data. Third, strain calculations in this study were performed by assuming the fully open aortic valve configuration as the reference, strain-free state and quantifying deformation between the open and closed extremes of the valve deformation. Although this assumption provides a practical framework for quantifying relative strain patterns associated with valve closure, it neglects the presence of residual, shear, and bending strains known to exist in vivo during systole and throughout leaflet coaptation. Consequently, the reported strain values should be interpreted as relative measures rather than absolute physiological strains. Finally, this study focused primarily on geometric and strain-based analyses without direct correlation to long-term clinical outcomes or functional measures. Future studies are therefore needed to establish relationships between the observed mechanical patterns and clinically relevant measures. Nonetheless, the proposed framework provides an important foundation for such investigations.

Importantly, the detailed strain distributions obtained through the proposed method can provide potential clinical insights: regions of abnormally high or low strain may indicate a higher risk of tissue remodeling, dysfunction, or failure, which could guide repair strategies and risk assessment. Therefore, this framework supports the development of individualized therapeutic strategies and advances our understanding of patient-specific valve behavior.

\vspace{0.5cm}
\sectionstylee{Author contribution} Writing – original draft: Mohsen Nakhaei; Methodology: Mohsen Nakhaei, Alison Pouch, Matthew Jolley, Wensi Wu; Formal analysis: Mohsen Nakhaei; Data curation: Silvani Amin,  Matthew Daemer, Christian Herz, Natalie Yushkevich, Lourdes Al Ghofaily, Nimesh Desai, Joseph Bavaria; Writing – review \& editing: Alison Pouch, Silvani Amin,  Matthew Daemer, Christian Herz, Natalie Yushkevich, Lourdes Al Ghofaily, Nimesh Desai, Joseph Bavaria, Matthew Jolley, Wensi Wu; Supervision: Alison Pouch, Matthew Jolley, Wensi Wu.

\vspace{0.5cm}
\sectionstylee{Funding}   This work was supported by the Cora Topolewski Pediatric Valve Center at the Children's Hospital of Philadelphia, Cardiac Center Innovation Grant, an Additional Ventures Single Ventricle Research Fund award, and the National Institutes of Health NHLBI K25 HL168235 (WW), NHLBI R01 HL153166 (MAJ), NHLBI R01 HL163202 (AMP).

\vspace{0.5cm}
\sectionstylee{Data availability} The data will be available upon request to the corresponding authors.
\vspace{0.5cm}
\section{Declarations}
\textbf{Conflicts of Interest} The authors declare that they have no competing interests or personal relationships that could have appeared to influence
the work reported in this paper.


\bibliographystyle{elsarticle-num}

\bibliography{references.bib}

@article{wu2018anisotropic,
  title     = {An anisotropic constitutive model for immersogeometric fluid--structure interaction analysis of bioprosthetic heart valves},
  author    = {Wu, Michael C. H. and Zakerzadeh, Rana and Kamensky, David and Kiendl, Josef and Sacks, Michael S. and Hsu, Ming-Chen},
  journal   = {Journal of Biomechanics},
  volume    = {74},
  pages     = {23--31},
  year      = {2018},
  publisher = {Elsevier},
  doi       = {10.1016/j.jbiomech.2018.04.012}
}

@article{sun2005finite,
  title     = {Finite element implementation of a generalized Fung-elastic constitutive model for planar soft tissues},
  author    = {Sun, Wei and Sacks, Michael S.},
  journal   = {Biomechanics and Modeling in Mechanobiology},
  volume    = {4},
  pages     = {190--199},
  year      = {2005},
  publisher = {Springer},
  doi       = {10.1007/s10237-005-0075-x}
}

@article{lee2014inverse,
  title     = {An inverse modeling approach for stress estimation in mitral valve anterior leaflet valvuloplasty for in-vivo valvular biomaterial assessment},
  author    = {Lee, Chung-Hao and Amini, Rouzbeh and Gorman, Robert C. and Gorman, Joseph H. III and Sacks, Michael S.},
  journal   = {Journal of Biomechanics},
  volume    = {47},
  number    = {9},
  pages     = {2055--2063},
  year      = {2014},
  publisher = {Elsevier},
  doi       = {10.1016/j.jbiomech.2013.10.058}
}

@article{fan2014simulation,
  title     = {Simulation of planar soft tissues using a structural constitutive model: finite element implementation and validation},
  author    = {Fan, Rong and Sacks, Michael S.},
  journal   = {Journal of Biomechanics},
  volume    = {47},
  number    = {9},
  pages     = {2043--2054},
  year      = {2014},
  publisher = {Elsevier},
  doi       = {10.1016/j.jbiomech.2014.03.014}
}

@article{kamensky2018contact,
  title     = {A contact formulation based on a volumetric potential: Application to isogeometric simulations of atrioventricular valves},
  author    = {Kamensky, David and Xu, Fei and Lee, Chung-Hao and Yan, Jinhui and Bazilevs, Yuri and Hsu, Ming-Chen},
  journal   = {Computer Methods in Applied Mechanics and Engineering},
  volume    = {330},
  pages     = {522--546},
  year      = {2018},
  publisher = {Elsevier},
  doi       = {10.1016/j.cma.2017.11.007}
}

@article{armfield2024effect,
  title     = {Effect of bioprosthetic leaflet anisotropy on stent dynamics of Transcatheter Aortic Valve Replacement devices},
  author    = {Armfield, Dylan and Boxwell, Sam and McNamara, Laoise and Cook, Scott and Conway, Shane and Celikin, Mert and Cardiff, Philip},
  journal   = {Journal of the Mechanical Behavior of Biomedical Materials},
  volume    = {157},
  pages     = {106650},
  year      = {2024},
  publisher = {Elsevier},
  doi       = {10.1016/j.jmbbm.2024.106650}
}

@article{khalili2022transvalvular,
  title     = {Transvalvular pressure gradients and all-cause mortality following TAVR: a multicenter echocardiographic and invasive registry},
  author    = {Khalili, Houman and Pibarot, Philippe and Hahn, Rebecca T. and Elmariah, Sammy and Pilgrim, Thomas and Bavry, Anthony A. and Maini, Brijeshwar and Okuno, Taishi and Al-Azizi, Karim and Waggoner, Thomas E. and others},
  journal   = {JACC: Cardiovascular Interventions},
  volume    = {15},
  number    = {18},
  pages     = {1837--1848},
  year      = {2022},
  publisher = {American College of Cardiology},
  doi       = {10.1016/j.jcin.2022.07.033}
}

@article{aggarwal2016vivo,
  title     = {In-vivo heterogeneous functional and residual strains in human aortic valve leaflets},
  author    = {Aggarwal, Ankush and Pouch, Alison M. and Lai, Eric and Lesicko, John and Yushkevich, Paul A. and Gorman, Joseph H. III and Gorman, Robert C. and Sacks, Michael S.},
  journal   = {Journal of Biomechanics},
  volume    = {49},
  number    = {12},
  pages     = {2481--2490},
  year      = {2016},
  publisher = {Elsevier},
  doi       = {10.1016/j.jbiomech.2016.04.038}
}

@article{rego2022patient,
  title     = {Patient-specific quantification of normal and bicuspid aortic valve leaflet deformations from clinically derived images},
  author    = {Rego, Bruno V. and Pouch, Alison M. and Gorman, Joseph H. III and Gorman, Robert C. and Sacks, Michael S.},
  journal   = {Annals of Biomedical Engineering},
  volume    = {50},
  number    = {1},
  pages     = {1--15},
  year      = {2022},
  publisher = {Springer},
  doi       = {10.1007/s10439-021-02882-0}
}

@article{abbasi2015leaflet,
  title     = {Leaflet stress and strain distributions following incomplete transcatheter aortic valve expansion},
  author    = {Abbasi, Mostafa and Azadani, Ali N.},
  journal   = {Journal of Biomechanics},
  volume    = {48},
  number    = {13},
  pages     = {3663--3671},
  year      = {2015},
  publisher = {Elsevier},
  doi       = {10.1016/j.jbiomech.2015.08.012}
}

@article{marom2012fluid,
  title     = {A fluid--structure interaction model of the aortic valve with coaptation and compliant aortic root},
  author    = {Marom, Gil and Haj-Ali, Rami and Raanani, Ehud and Sch{\"a}fers, Hans-Joachim and Rosenfeld, Moshe},
  journal   = {Medical \& Biological Engineering \& Computing},
  volume    = {50},
  number    = {2},
  pages     = {173--182},
  year      = {2012},
  publisher = {Springer},
  doi       = {10.1007/s11517-011-0849-5}
}

@article{jermihov2011effect,
  title     = {Effect of geometry on the leaflet stresses in simulated models of congenital bicuspid aortic valves},
  author    = {Jermihov, Paul N. and Jia, Lu and Sacks, Michael S. and Gorman, Robert C. and Gorman, Joseph H. III and Chandran, Krishnan B.},
  journal   = {Cardiovascular Engineering and Technology},
  volume    = {2},
  number    = {1},
  pages     = {48--56},
  year      = {2011},
  publisher = {Springer},
  doi       = {10.1007/s13239-011-0035-9}
}

@article{ayoub2017regulation,
  title     = {Regulation of valve interstitial cell homeostasis by mechanical deformation: implications for heart valve disease and surgical repair},
  author    = {Ayoub, Salma and Lee, Chung-Hao and Driesbaugh, Kathryn H. and Anselmo, Wanda and Hughes, Connor T. and Ferrari, Giovanni and Gorman, Robert C. and Gorman, Joseph H. and Sacks, Michael S.},
  journal   = {Journal of the Royal Society Interface},
  volume    = {14},
  number    = {135},
  pages     = {20170580},
  year      = {2017},
  publisher = {The Royal Society},
  doi       = {10.1098/rsif.2017.0580}
}

@article{py06nimg,
  title     = {User-guided {3D} active contour segmentation of anatomical structures: significantly improved efficiency and reliability},
  author    = {Yushkevich, Paul A. and Piven, Joseph and Hazlett, Heather Cody and Smith, Rachel Gimpel and Ho, Sean and Gee, James C. and Gerig, Guido},
  journal   = {NeuroImage},
  volume    = {31},
  number    = {3},
  pages     = {1116--1128},
  year      = {2006},
  publisher = {Elsevier},
  doi       = {10.1016/j.neuroimage.2006.01.015}
}

@article{maas2012febio,
  title     = {{FEBio}: Finite elements for biomechanics},
  author    = {Maas, Steve A. and Ellis, Benjamin J. and Ateshian, Gerard A. and Weiss, Jeffrey A.},
  journal   = {Journal of Biomechanical Engineering},
  volume    = {134},
  number    = {1},
  pages     = {011005},
  year      = {2012},
  publisher = {ASME},
  doi       = {10.1115/1.4005694}
}

@article{fedorov20123d,
  title     = {{3D Slicer} as an image computing platform for the Quantitative Imaging Network},
  author    = {Fedorov, Andriy and Beichel, Reinhard and Kalpathy-Cramer, Jayashree and Finet, Julien and Fillion-Robin, Jean-Christophe and Pujol, Sonia and Bauer, Christian and Jennings, Dominique and Fennessy, Fiona and Sonka, Milan and others},
  journal   = {Magnetic Resonance Imaging},
  volume    = {30},
  number    = {9},
  pages     = {1323--1341},
  year      = {2012},
  publisher = {Elsevier},
  doi       = {10.1016/j.mri.2012.05.001}
}

@article{herz2024open,
  title     = {Open-source graphical user interface for the creation of synthetic skeletons for medical image analysis},
  author    = {Herz, Christian and Vergnet, Nicolas and Tian, Sijie and Aly, Abdullah H. and Jolley, Matthew A. and Tran, Nathanael and Arenas, Gabriel and Lasso, Andras and Schwartz, Nadav and O'Neill, Kathleen E. and others},
  journal   = {Journal of Medical Imaging},
  volume    = {11},
  number    = {3},
  pages     = {036001},
  year      = {2024},
  publisher = {SPIE},
  doi       = {10.1117/1.JMI.11.3.036001}
}

@article{sahasakul1988age,
  title     = {Age-related changes in aortic and mitral valve thickness: implications for two-dimensional echocardiography based on an autopsy study of 200 normal human hearts},
  author    = {Sahasakul, Yongyuth and Edwards, William D. and Naessens, James M. and Tajik, A. Jamil},
  journal   = {The American Journal of Cardiology},
  volume    = {62},
  number    = {7},
  pages     = {424--430},
  year      = {1988},
  publisher = {Elsevier},
  doi       = {10.1016/0002-9149(88)90971-x}
}

@article{huang2007situ,
  title     = {In-situ deformation of the aortic valve interstitial cell nucleus under diastolic loading},
  author    = {Huang, Hsiao-Ying Shadow and Liao, Jun and Sacks, Michael S.},
  journal   = {Journal of Biomechanical Engineering},
  volume    = {129},
  number    = {6},
  pages     = {880--889},
  year      = {2007},
  publisher = {ASME},
  doi       = {10.1115/1.2801670}
}

@article{evangelista2011bicuspid,
  title     = {Bicuspid aortic valve and aortic root disease},
  author    = {Evangelista, Artur},
  journal   = {Current Cardiology Reports},
  volume    = {13},
  number    = {3},
  pages     = {234--241},
  year      = {2011},
  publisher = {Springer},
  doi       = {10.1007/s11886-011-0175-4}
}

@article{merritt2014association,
  title     = {Association between leaflet fusion pattern and thoracic aorta morphology in patients with bicuspid aortic valve},
  author    = {Merritt, Bryce A. and Turin, Alexander and Markl, Michael and Malaisrie, S. Chris and McCarthy, Patrick M. and Carr, James C.},
  journal   = {Journal of Magnetic Resonance Imaging},
  volume    = {40},
  number    = {2},
  pages     = {294--300},
  year      = {2014},
  publisher = {Wiley},
  doi       = {10.1002/jmri.24376}
}

@article{davies1996demographic,
  title     = {Demographic characteristics of patients undergoing aortic valve replacement for stenosis: relation to valve morphology},
  author    = {Davies, M. J. and Treasure, T. and Parker, D. J.},
  journal   = {Heart},
  volume    = {75},
  number    = {2},
  pages     = {174--178},
  year      = {1996},
  publisher = {BMJ},
  doi       = {10.1136/hrt.75.2.174}
}

@article{yap2010dynamic,
  title     = {Dynamic hemodynamic energy loss in normal and stenosed aortic valves},
  author    = {Yap, Choon-Hwai and Dasi, Lakshmi P. and Yoganathan, Ajit P.},
  journal   = {Journal of Biomechanical Engineering},
  volume    = {132},
  number    = {2},
  pages     = {021002},
  year      = {2010},
  publisher = {ASME},
  doi       = {10.1115/1.4000874}
}

@article{ayoub2021role,
  title     = {On the role of predicted in vivo mitral valve interstitial cell deformation on its biosynthetic behavior},
  author    = {Ayoub, Salma and Howsmon, Daniel P. and Lee, Chung-Hao and Sacks, Michael S.},
  journal   = {Biomechanics and Modeling in Mechanobiology},
  volume    = {20},
  number    = {1},
  pages     = {135--144},
  year      = {2021},
  publisher = {Springer},
  doi       = {10.1007/s10237-020-01373-w}
}

@article{lam2016valve,
  title     = {Valve interstitial cell contractile strength and metabolic state are dependent on its shape},
  author    = {Lam, Ngoc Thien and Muldoon, Timothy J. and Quinn, Kyle P. and Rajaram, Narasimhan and Balachandran, Kartik},
  journal   = {Integrative Biology},
  volume    = {8},
  number    = {10},
  pages     = {1079--1089},
  year      = {2016},
  publisher = {Oxford University Press},
  doi       = {10.1039/c6ib00120c}
}

@article{decano2022disease,
  title     = {A disease-driver population within interstitial cells of human calcific aortic valves identified via single-cell and proteomic profiling},
  author    = {Decano, Julius L. and Iwamoto, Yukio and Goto, Shinji and Lee, Janey Y. and Matamalas, Joan T. and Halu, Arda and Blaser, Mark and Lee, Lang Ho and Pieper, Brett and Chelvanambi, Sarvesh and others},
  journal   = {Cell Reports},
  volume    = {39},
  number    = {2},
  pages     = {110685},
  year      = {2022},
  publisher = {Elsevier},
  doi       = {10.1016/j.celrep.2022.110685}
}

@article{aggarwal2014architectural,
  title     = {Architectural trends in the human normal and bicuspid aortic valve leaflet and its relevance to valve disease},
  author    = {Aggarwal, Ankush and Ferrari, Giovanni and Joyce, Erin and Daniels, Michael J. and Sainger, Rachana and Gorman, Joseph H. III and Gorman, Robert and Sacks, Michael S.},
  journal   = {Annals of Biomedical Engineering},
  volume    = {42},
  number    = {5},
  pages     = {986--998},
  year      = {2014},
  publisher = {Springer},
  doi       = {10.1007/s10439-014-0973-0}
}

@article{rutkovskiy2017valve,
  title     = {Valve interstitial cells: the key to understanding the pathophysiology of heart valve calcification},
  author    = {Rutkovskiy, Arkady and Malashicheva, Anna and Sullivan, Gareth and Bogdanova, Maria and Kostareva, Anna and Stensl{\o}kken, K{\aa}re-Olav and Fiane, Arnt and Vaage, Jarle},
  journal   = {Journal of the American Heart Association},
  volume    = {6},
  number    = {9},
  pages     = {e006339},
  year      = {2017},
  doi       = {10.1161/JAHA.117.006339}
}

@article{khang2023three,
  title     = {Three-dimensional analysis of hydrogel-imbedded aortic valve interstitial cell shape and its relation to contractile behavior},
  author    = {Khang, Alex and Nguyen, Quan and Feng, Xinzeng and Howsmon, Daniel P. and Sacks, Michael S.},
  journal   = {Acta Biomaterialia},
  volume    = {163},
  pages     = {194--209},
  year      = {2023},
  publisher = {Elsevier},
  doi       = {10.1016/j.actbio.2022.01.039}
}

@article{chen2022image,
  title     = {Image registration-based method for reconstructing transcatheter heart valve geometry from patient-specific CT scans},
  author    = {Chen, Huang and Yeats, Breandan and Swamy, Kevin and Samaee, Milad and Sivakumar, Sri Krishna and Esmailie, Fateme and Razavi, Atefeh and Yadav, Pradeep and Thourani, Vinod H. and Polsani, Venkateshwar and others},
  journal   = {Annals of Biomedical Engineering},
  volume    = {50},
  number    = {7},
  pages     = {805--815},
  year      = {2022},
  publisher = {Springer},
  doi       = {10.1007/s10439-022-02962-9}
}

@article{lior2023semi,
  title     = {Semi-automated construction of patient-specific aortic valves from computed tomography images},
  author    = {Lior, Dan and Puelz, Charles and Edwards, Colin and Molossi, Silvana and Griffith, Boyce E. and Birla, Ravi K. and Rusin, Craig G.},
  journal   = {Annals of Biomedical Engineering},
  volume    = {51},
  number    = {1},
  pages     = {189--199},
  year      = {2023},
  publisher = {Springer},
  doi       = {10.1007/s10439-022-03075-z}
}

@article{labrosse2015subject,
  title     = {Subject-specific finite-element modeling of normal aortic valve biomechanics from 3D+ t TEE images},
  author    = {Labrosse, Michel R. and Beller, Carsten J. and Boodhwani, Munir and Hudson, Christopher and Sohmer, Benjamin},
  journal   = {Medical Image Analysis},
  volume    = {20},
  number    = {1},
  pages     = {162--172},
  year      = {2015},
  publisher = {Elsevier},
  doi       = {10.1016/j.media.2014.11.003}
}

@article{yin2024fluid,
  title     = {Fluid--structure interaction analysis of a healthy aortic valve and its surrounding haemodynamics},
  author    = {Yin, Zhongjie and Armour, Chl{\"o}e and Kandail, Harkamaljot and O'Regan, Declan P. and Bahrami, Toufan and Mirsadraee, Saeed and Pirola, Selene and Xu, Xiao Yun},
  journal   = {International Journal for Numerical Methods in Biomedical Engineering},
  volume    = {40},
  number    = {11},
  pages     = {e3865},
  year      = {2024},
  publisher = {Wiley},
  doi       = {10.1002/cnm.3865}
}

@article{west2023effects,
  title     = {The effects of strain history on aortic valve interstitial cell activation in a 3D hydrogel environment},
  author    = {West, Toni M. and Howsmon, Daniel P. and Massidda, Miles W. and Vo, Helen N. and Janobas, Athena A. and Baker, Aaron B. and Sacks, Michael S.},
  journal   = {APL Bioengineering},
  volume    = {7},
  number    = {2},
  year      = {2023},
  publisher = {AIP Publishing},
  doi       = {10.1063/5.0138030}
}

@article{kim2008dynamic,
  title     = {Dynamic simulation of bioprosthetic heart valves using a stress resultant shell model},
  author    = {Kim, Hyunggun and Lu, Jia and Sacks, Michael S. and Chandran, Krishnan B.},
  journal   = {Annals of Biomedical Engineering},
  volume    = {36},
  number    = {2},
  pages     = {262--275},
  year      = {2008},
  publisher = {Springer},
  doi       = {10.1007/s10439-007-9409-4}
}

@book{hole1996hole,
  title     = {Hole's Human Anatomy \& Physiology},
  author    = {Hole, John and Shier, David and Butler, Jackie and Lewis, Ricki},
  year      = {1996},
  publisher = {WC Brown}
}

@article{yushkevich2016ic,
  title     = {IC-P-174: fast automatic segmentation of hippocampal subfields and medial temporal lobe subregions in 3 Tesla and 7 Tesla T2-Weighted MRI},
  author    = {Yushkevich, Paul A. and Pluta, John and Wang, Hongzhi and Wisse, Laura EM and Das, Sandhitsu and Wolk, David},
  journal   = {Alzheimer's \& Dementia},
  volume    = {12},
  pages     = {P126--P127},
  year      = {2016},
  publisher = {Wiley},
  doi       = {10.1016/j.jalz.2016.06.205}
}

@article{meredith2025aortic,
  title     = {Aortic valve leaflet motion for diagnosis and classification of aortic stenosis using single view echocardiography},
  author    = {Meredith, Thomas and Mohammed, Farhan and Pomeroy, Amy and Barbieri, Sebastiano and Meijering, Erik and Jorm, Louisa and Roy, David and Hayward, Christopher and Kovacic, Jason C. and Muller, David WM and others},
  journal   = {Journal of Cardiovascular Imaging},
  volume    = {33},
  number    = {1},
  pages     = {8},
  year      = {2025},
  publisher = {Springer},
  doi       = {10.1186/s44348-025-00051-8}
}

@article{Yang2020_AR_DBP_RHR,
  title     = {Diastolic Blood Pressure and Heart Rate Are Independently Associated With Mortality in Chronic Aortic Regurgitation},
  author    = {Yang, Li-Tan and Pellikka, Patricia A. and Enriquez-Sarano, Maurice and Scott, Christopher G. and Padang, Ratnasari and Mankad, Sunil V. and Schaff, Hartzell V. and Michelena, Hector I.},
  journal   = {Journal of the American College of Cardiology},
  volume    = {75},
  number    = {1},
  pages     = {29--39},
  year      = {2020},
  doi       = {10.1016/j.jacc.2019.10.047}
}

@inproceedings{wu2024physics,
  title     = {Physics in the loop: Integrating biomechanics-derived training data into a neural ordinary differential equation-based deformable registration framework},
  author    = {Wu, Wensi and Wu, Yifan and Sulentic, Analise M. and Gee, James and Pouch, Alison Marie and Jolley, Matthew A.},
  booktitle = {Medical Imaging with Deep Learning},
  year      = {2024}
}

@article{Lasso2022_SlicerHeart,
  title     = {SlicerHeart: An Open-Source Computing Platform for Cardiac Image Analysis and Modeling},
  author    = {Lasso, Andras and Herz, Christian and Nam, Hannah and Cianciulli, Alana and Pieper, Steve and Drouin, Simon and Pinter, Csaba and St-Onge, Samuelle and Vigil, Chad and Ching, Stephen and Sunderland, Kyle and Fichtinger, Gabor and Kikinis, Ron and Jolley, Matthew A.},
  journal   = {Frontiers in Cardiovascular Medicine},
  volume    = {9},
  pages     = {886549},
  year      = {2022},
  doi       = {10.3389/fcvm.2022.886549}
}

\end{document}